\documentclass[aos,preprint]{imsart}
\usepackage{mathrsfs}
\usepackage{graphicx}
\usepackage{amsmath}
\usepackage{color}
\usepackage{rotating}
\usepackage{lscape}
\usepackage{graphicx}
\usepackage{bm}
\usepackage{mathrsfs}
\usepackage{booktabs}
\usepackage{caption}
\usepackage{subfig}
\usepackage{float}
\usepackage{amsfonts}

\def\be{\begin{equation}}
\def\ee{\end{equation}}
\def\bea{\begin{eqnarray}}
\def\eea{\end{eqnarray}}
\def\bd{\begin{displaymath}}
\def\ed{\end{displaymath}}
\def\bda{\begin{eqnarray*}}
\def\eda{\end{eqnarray*}}
\def\bsm{\begin{small}}
\def\esm{\end{small}}

\def\ha1{\hat \beta_1}

\def\bb0{\delta_\beta}

\def\be{\begin{equation}}
\def\ee{\end{equation}}
\def\bea{\begin{eqnarray}}
\def\eea{\end{eqnarray}}
\def\bd{\begin{displaymath}}
\def\ed{\end{displaymath}}
\def\bda{\begin{eqnarray*}}
\def\eda{\end{eqnarray*}}
\def\bsm{\begin{small}}
\def\esm{\end{small}}
\newcommand{\E}{\rm E}
\newcommand{\V}{\rm Var}

\def\bsc{\begin{scriptsize}}
\def\esc{\end{scriptsize}}

\def\qed{\qquad$\Box$\medskip}
\input{epsf.sty}
\usepackage{epsfig}

\RequirePackage[OT1]{fontenc}
\RequirePackage{amsthm,amsmath}
\RequirePackage[numbers]{natbib}
\RequirePackage[colorlinks,citecolor=blue,urlcolor=blue]{hyperref}
\startlocaldefs
\numberwithin{equation}{section}
\theoremstyle{plain}

\endlocaldefs
\setattribute{journal}{name}{}

\begin{document}
\begin{frontmatter}

\title{Online Change-Point Detection in High-Dimensional Covariance Structure with application to dynamic networks}
\runtitle{High-Dimensional Online Change-Point Detection}

\begin{aug}
\author{\fnms{Lingjun} \snm{Li}\thanksref{m1}\ead[label=e1]{lli13@kent.edu }}
\and
\author{\fnms{Jun} \snm{Li}\thanksref{m2}\ead[label=e2]{jli49@kent.edu}}
\affiliation{Kent State University\thanksmark{m1} and Kent State University\thanksmark{m2}}

\address{Department of Mathematical Sciences\\
Kent State University\\
Kent,Ohio 44242,
USA\\
\printead{e1}\\
\phantom{E-mail:\ }}

\address{Department of Mathematical Sciences\\
Kent State University\\
Kent,Ohio 44242,
USA\\
\printead{e2}}

\end{aug}

\begin{abstract}
%Different from offline data, online data are the real-time observations which continually arrive. 
%One important task in online data analysis is detecting network change, such as dissociation of communities or formation of new communities. Targeting on this type of application,
In this paper, we develop an online change-point detection procedure in the covariance structure of high-dimensional data. A new stopping rule is proposed to terminate the process as early as possible when a change in covariance structure occurs. The stopping rule allows temporal dependence and can be applied to non-Gaussian data. An explicit expression for the average run length (ARL) is derived, so that the level of threshold in the stopping rule can be easily obtained with no need to run time-consuming Monte Carlo simulations. We also establish an upper bound for the expected detection delay (EDD), the expression of which demonstrates the impact of data dependence and magnitude of change in the covariance structure.
Simulation studies are provided to confirm accuracy of the theoretical results. The practical usefulness of the proposed procedure is illustrated by detecting the change of brain's covariance network in a resting-state fMRI dataset.
\end{abstract}

\begin{keyword}[class=MSC]
\kwd[Primary ]{}
\kwd{62G20}
\kwd[; secondary ]{}
\kwd{62G10}
\end{keyword}
\begin{keyword}
\kwd{Change-point detection}
\kwd{High-dimensional data}
\end{keyword}

\end{frontmatter}

%\vspace*{.1in}

%\noindent\textsc{Keywords}: {Hotelling's $T^2$ test; Large $p$ small $n$. }

%\newpage

\section{Introduction}

Online change-point detection or sequential change-point detection, originally arises from the problem of quality control. The product quality is monitored based on the observations continually arriving during an industrial process.  A stopping rule is chosen to terminate and reset the process as early as possible when an anomaly occurs.
In modern applications, there has been a resurgence of interest in detecting abrupt change from streaming data with a large number of measurements. %to the areas including finance, clinical trials and cybersercurity.
Examples include real-time monitoring for sensor networks and threat detection from surveillance videos. More can be found in studying dynamic connectivity of resting state functional magnetic resonance imaging, and in detecting threat of fake news from the group of fake accounts in social networks (Bara, Fung and Dinh 2015). 

Extensive research has been done for online change-point detection of univariate data; see, for example, Page (1954), Shiryayev (1963), Lorden (1971), Wald (1973), Siegmund (1985) and Siegmund and Venkatraman (1995). The proposed stopping rules are based on the CUSUM test or the quasi-Bayesian test which assume the distributions of data before and after the change point to be known, or its variants proposed to relax the restrictive assumption of known distributions.
There also exist many developments in online change-point detection of multivariate data. For example, Tartakovsky and Veeravalli (2008) and Mei (2010) propose the stopping rule for the common change point detection from all dimensions based on the assumption that the distributions of data before and after the change point are known. By relaxing the common change to the change of only subset of data, Xie and Siegmund (2013), and Chan and Walther (2015) study the stopping rule for the multivariate normally distributed data with the identity covariance matrix. By extending and modifying the approach in Xie and Siegmund (2013), Chan (2017) investigates the optimal detection of multiple data streams in detecting mean shift of independent multivariate normally distributed data with the identity covariance matrix. Despite the preceding developments, very little work has been done for online change-point detection of high-dimensional data. A recent development can be seen in Chen (2019) where the proposed stopping rule utilizes nearest neighbor information to detect the change point in the distribution of independent data.     %In the era of big data, high-dimensional sequential data have been commonly seen in many areas including finance, clinical trials and cybersercurity.

%Despite the aforementioned achievements, they %for multivariate sequential data

In this paper, we consider online change-point detection in the covariance structure of high-dimensional data. More precisely, letting $\{X_1, X_2, \cdots \}$ be a sequence of continually arriving $p$-dimensional random vectors, each of which has its own covariance matrix $\Sigma_i$, we consider the hypotheses %whether $\Sigma_1=\cdots=\Sigma_{\tau}\ne \Sigma_{\tau+1}=\cdots$ at some unknown time point $\tau$.
\begin{eqnarray}
&H_0&: \Sigma_1=\Sigma_2=\cdots, \qquad \mbox{against} \nonumber\\
&H_1&: \Sigma_1=\cdots=\Sigma_{\tau} \ne \Sigma_{\tau+1}=\cdots, \label{Hypo}
\end{eqnarray}
where $\tau$ is an unknown change point. The motivation behind the considered problem stems from the applications of detecting covariance network changes, such as dynamic changes in brain's functional connectivity, where the network can be quantified by the covariance or precision (inverse covariance) matrix (Varoquaux et al., 2010). Since our main focus is change point detection of covariance structure, the population mean is assumed to be constant to facilitate our analysis. We propose a stopping rule for (\ref{Hypo}), which terminates the process as early as possible after $\Sigma_{\tau}$ changes to $\Sigma_{\tau+1}$. Under the null hypothesis, we derive an explicit expression for the average run length (ARL), so that the level of threshold in the stopping rule can be easily obtained with no need to run time-consuming Monte Carlo simulations. Under the alternative hypothesis, we establish an upper bound for the expected detection delay (EDD), which demonstrates the impact of data dependence and magnitude of change in the covariance structure.

The proposed stopping rule is readily applied to detecting covariance network change in high-dimensional online data. %as the network can be modeled by the covariance matrix. %For example, fake accounts in social networks such as Facebook, Twitter and Instagram, tend to follow similar patterns when posting fake messages, hurtful photographs and unpleasant comments (Bara, Fung and Dinh 2015). It is thus more efficient to detect threat of fake news based on the group of fake accounts rather than the individual accounts. %We will propose a nonparametric sequential testing procedure for the above hypotheses.
In addition to its practical usefulness, the developed method has several theoretical contributions. First, unlike many other detection procedures that assume temporal independence, the proposed stopping rule %incorporates spatial dependence among the components of each high-dimensional measurement at each particular time point, and 
allows temporal dependence among different high-dimensional measurements at different time points. %In order to control the ARL at any pre-specified value, the distribution of the stopping time needs to be established as a function of the threshold. If the distribution under the assumption of no temporal dependence does hold when the temporal dependence is present, it is more likely to generate a false alarm which terminates the process earlier than expected. 
Moreover, %rather than assume temporal independence, 
we estimate the temporal dependence consistently through a data-driven procedure, and establish the distribution of the stopping time with the correctly specified dependence. Consequently, the ARL of the proposed stopping rule can be well controlled even in the presence of temporal dependence.      
Second, the stopping rule can be applied to a wide range of data in that it does not assume Gaussian distribution, but only requires existence of fourth moment of data. 
Third, the stopping rule is implementable when the dimension $p$ diverges and thus suitable for monitoring modern networks whose size varies enormously from thousands to millions.  
Finally, we identify the key factors and establish their impact on the EDD through an explicitly derived upper bound. In particular, we reveal that the EDD based on the $L_2$-norm statistic increases as the strength of temporal dependence increases, but decreases as the magnitude of change $||\Sigma_{\tau+1}-\Sigma_{\tau}||_F/||\Sigma_{\tau}||_F$ increases. Here $||\cdot||_F$ represents the matrix Frobenius norm. %This reveals that the proposed stopping rule is efficient in detecting the change when the two covariance matrices differ in a large number of coordinates. When they only differ in a small number of coordinates, the value of $||\Sigma_{\tau+1}-\Sigma_{\tau}||_F$ can be diluted by $||\Sigma_{\tau}||_F$. Our result suggests that a future research for this scenario may consider excluding the coordinates without covariance change and improve the EDD by reducing $||\Sigma_{\tau}||_F$ to the Frobenius norm of the covariance matrix of the remaining coordinates.    

The rest of the paper is organized as follows. Section 2 introduces the proposed stopping rule. Section 3 presents its asymptotic properties. Simulation studies and real data analysis are given in Sections 4 and 5, respectively. We conclude the paper in Section 6. 
Technical proofs of main theorems are relegated to Appendix. Other technical proofs and additional simulation results are included in a supplementary material.

\section{Methodology}

%Rather than impose temporal independence, we allow both spatial and temporal dependence in high-dimensional online data. %an effective stopping rule should be constructed to incorporate such complex dependence.
%we first propose a non-parametric model to incorporate such complex dependence.

\subsection{Modeling spatial and temporal dependence}

Let $\{X_i, 1\le i \le n\}$ be a sequence of $p$-dimensional random vectors. As discussed in Section 1, we assume $\mbox{E}(X_i)=\mu$ to facilitate our analysis. %For the sake of simplicity, we assume the sequence has zero mean, %namely $\mbox{E}(X_i)=0$ for $i=1, \cdots, n$. Otherwise, each observation can be centralized by subtracting the sample mean.
We model the sequence by
\be
X_{i}=\mu+\Gamma_i Z \qquad \mbox{for} \quad i=1, \cdots, n,   \label{model}
\ee
where %$\mu_i$ is the $p$-dimensional population mean vector,
$\Gamma_i$ is a $p \times m$ matrix with $m\ge n\cdot p$, and $Z=(z_{1}, \cdots, z_{m})^{T}$ such that $\{z_i\}_{i=1}^m$ are mutually independent and satisfy $\mbox{E}(z_i)=0$, $\mbox{Var}(z_i)=1$ and $\mbox{E}(z_i^4)=3+\beta$ for some finite constant $\beta$. %requiring $\mbox{E}(Z)=0$, $\mbox{Var} (Z)= I_m$, the $m\times m$ identity matrix, and .  %and let $\Delta$ be a finite constant, then
%\begin{eqnarray}
%\mbox{E}(z_{k}^4)=3+\Delta, \quad \mbox{and} \quad \mbox{E}(z_{k_1}^{l_1}z_{k_2}^{l_2}\cdots z_{k_h}^{l_h})=\mbox{E}(z_{k_1}^{l_1})\mbox{E}(z_{k_2}^{l_2})\cdots \mbox{E}(z_{k_h}^{l_h}),\nonumber
%\end{eqnarray}
%where $h$ is positive integer such that $\sum_{j=1}^h l_h \le 8$ and $l_1\ne l_2 \ne \cdots \ne l_h$.

There are two advantages to impose the above model. First, it incorporates both spatial and temporal dependence of the sequence $\{X_i, 1\le i \le n\}$. Let $X=(X_1^T, \cdots, X_n^T)^T$ and $\Gamma=(\Gamma_1^T, \cdots, \Gamma_n^T)^T$. From (\ref{model}), the variance-covariance matrix of $X$ is $\Gamma \Gamma^T$, in which each $p \times p$ block diagonal sub-matrix $\Gamma_i\Gamma_i^T\equiv \Sigma_i$ represents the spatial dependence of each $X_i$ and each $p \times p$ block off-diagonal sub-matrix $\Gamma_i\Gamma_j^T \equiv C(j-i)$ describes the spatial and temporal dependence between $X_i$ and $X_j$ at $i \ne j$. Here we require $m \ge n\times p$ to ensure the positive definite of $\Gamma \Gamma^T$ and thus existence of $C(j-i)$. Second, the model does not assume any distribution of data, but only requires the existence of fourth moment. In particular, $X_i$ is normally distributed if $\beta=0$. 

Based on (\ref{model}), we accommodate the spatial and temporal dependence by the following two conditions.

\medskip
(C1). The sequence is $M$-dependent, such that for some integer $M \ge 0$, $C(j-i) \ne 0$ if and only if $|j-i| \le M$. Moreover, under $H_0$ of (\ref{Hypo}), $C(j-i)=C(h)$ for all $i$ and $j$ satisfying $j-i=h$ with $h \in \{0, \pm 1, \cdots, \pm M\}$.

\medskip
Under the null hypothesis, we assume that the sequence is $M$-dependent, and the spatial and temporal dependence is stationary.  Under the alternative hypothesis, the covariance structure changes and consequently, the stationary of the spatial and temporal dependence cannot hold. We thus only assume the $M$-dependence.  %We thus only assume the  under the alternative. %We do not assume the stationary under the alternative hypothesis because it is not desirable that the sequence still retains the stationary if its covariance encounters any structural change. Otherwise, additional conditions are needed such that $\Gamma_i \Gamma_i^T=\Gamma_j \Gamma_j^T \ne \Gamma_k \Gamma_k^T=\Gamma_l \Gamma_l^T$ and $\Gamma_i \Gamma_j^T=\Gamma_k \Gamma_l^T$ for any $i, j$ before a change point and $k, l$ after the change point. This will make the model (\ref{model}) more restrictive. In order to provide insights into the performance of the proposed method, we first assume $M$ to be known. A data-driven procedure for choosing $M$ and extended theoretical results based on the estimated $M$ will be postponed to Section 3.3.
We introduce the $M$-dependence to relax the commonly assumed temporal independence in the literature. As shown in Appendix, the assumption enables us to establish the asymptotic normality of the test statistic (\ref{estimator}) through the martingale central limit theorem. Moreover, the $M$-dependence combined with the stationary in the spatial and temporal dependence, yields that the stopping time (\ref{stopping-rule}) converges to the Gumbel limiting distribution of a stationary Gaussian process under the null hypothesis. %The $M$-dependence can be generalized to the Assumption 1 in Li et al. (2019), where the spatial and temporal dependence is relatively weak when the timespan exceeds a critical value $M$.      
Under the alternative hypothesis, we impose the $M$-dependence to generalize the Wald's lemma from a sum of a random number of independent random variables to that of $M$-dependent random variables. The generalization enables us to study the EDD of the stopping time even in the presence of temporal dependence (see the proof of Theorem 2 in Appendix).  

\medskip
(C2). For any $h_1, h_2, h_3, h_4 \in \{0, \pm 1, \cdots, \pm M\}$, as $p \to \infty$, 
\[
\mbox{tr}\{C(h_1) C(h_2) C(h_3) C(h_4)\}=o\biggl[ \mbox{tr}\{C(h^{\prime}_1) C(h^{\prime}_2)\} \mbox{tr}\{C(h^{\prime}_3) C(h^{\prime}_4)\}\biggr],
\]
where $\{h^{\prime}_1, h^{\prime}_2, h^{\prime}_3, h^{\prime}_4\}$ is a permutation of $\{h_1, h_2, h_3, h_4\}$.

\medskip
%With a sequence of independent random vectors ($M=0$), (C2) is reduced to the conditions used in Chen and Qin (2010) and Li and Chen (2012) for two sample testing problem. Except the spatial dependence, the current condition is imposed to account for additional temporal dependence.
Without temporal dependence, (C2) becomes $\mbox{tr}\{C^4(0)\}=o[\mbox{tr}^2\{C^2(0)\}]$. %which is assumed in Chen and Qin (2010) and Li and Chen (2012) for the two sample testing problems. 
It holds if all the eigenvalues of $C(0)$ are bounded, but violates under strong dependence such as the compound symmetry covariance structure. If the temporal dependence is present ($h \ne 0$), (C2) takes into account both spatial and temporal dependence. It can be shown that (C2) holds if the requirement of bounded eigenvalues is extended to the $np \times np$ covariance matrix of entire sequence $X=(X_1^{T}, X_2^{T}, \cdots, X_n^{T})^{T}$, each $p \times p$ block diagonal matrix of which measures the spatial dependence of each $p$-dimensional random vector in the sequence, and each $p \times p$ block off-diagonal matrix of which describes the spatio-temporal dependence of two random vectors collected at different time points. The condition cannot hold if the spatial and temporal dependence is too strong so that the covariance matrix of $X$ has unbounded eigenvalues. %The condition excludes strong dependence case such as the equal correlation structure in which one eigenvalue diverges as $p$ goes to infinity, so that $\mbox{tr}\{C^4(0)\}=o[\mbox{tr}^2\{C^2(0)\}]$ does not hold.             
The advantage of (C2) is that it does not impose any decay structures on $C(h)$ as long as the trace condition is satisfied. Moreover, it allows the dimension $p$ to diverge without imposing its growth rate. 
%We combine (C2) with (\ref{model}) and (C1) to establish the asymptotic distributions of the test statistic (\ref{estimator}) and the stopping time (\ref{stopping-rule}), without imposing any explicit restriction on the data dimension $p$.  This makes the stopping rule suitable for modern network analysis in which data size can vary from thousands to millions.     

\subsection{Test statistic}

%A stopping rule needs to detect the change point as soon as possible when an anomaly occurs.
%For the online change-point detection problem (\ref{Hypo}), 
Suppose that $n$ observations have been collected. We need a test statistic, the expectation of which can measure the heterogeneity of covariance structure from the collected observations. Assuming for the moment that $\mu=0$ in (\ref{model}), we propose the following test statistic
\be
\hat{\mathcal{J}}_{n, M}\equiv \frac{1}{n^2} \sum_{i, j=1}^n  W_{M}(i, j)(X_i^T X_j)^2, \label{estimator}
\ee
where the weight function $W_{M}(i, j) \equiv \sum_{t=M+2}^{n-M-2}A_{t, M}(i, j)\mbox{I}(|i-j|\ge M+1)$ and %$\mathcal{A}(i, j)=\sum_{t=M+2}^{n-M-2} A_t(i, j)\mbox{I}(|i-j|\ge M+1)$ and
\begin{eqnarray}
&\quad&A_{t, M}(i, j)\nonumber\\
&=&\frac{n-t-M}{t-M-1}\mbox{I}(i \le t)\mbox{I}(j \le t) + \frac{t-M}{n-t-M-1}\mbox{I}(t+1 \le i)\mbox{I}(t+1 \le j) \nonumber\\
&-&\frac{(t-M)(n-t-M)}{t(n-t)-\frac{1}{2}M(M+1)}\{\mbox{I}(i \le t) \mbox{I}(t+1 \le j )+\mbox{I}(t+1 \le i) \mbox{I}(j \le t)\}. \nonumber
\end{eqnarray}
If $\mu \ne 0$, a centralized version of (\ref{estimator}) is
\begin{eqnarray}
\hat{\mathcal{J}}^*_{n, M}=\frac{1}{n^2} \sum_{i, j=1}^n  W_{M}(i, j) \{(X_i-\hat{\mu})^T (X_j-\hat{\mu})\}^2, \label{estimator-cen}
\end{eqnarray}
where $\hat{\mu}$ is a consistent estimator of $\mu$. As introduced in Section 2.3, the proposed stopping rule needs a training sample and $\hat{\mu}$ thus can be chosen as the sample mean of the training sample. 

%We provide the following remarks to understand (\ref{estimator}).
\bigskip
{\bf Remark 2.1} We first assume a known $M$ to present the main results of the proposed methods. We then provide a data-driven procedure for estimating $M$ and establish the theoretical results based on the estimated $M$ in Section 3.4. 

\bigskip
{\bf Remark 2.2} The test statistics are constructed in several steps. At each $t$ from $\{M+2,\cdots, n-M-2\}$, we first partition the entire sequence $\{X_i, 1\le i \le n\}$ into two segments $\{X_i, 1\le i \le t \}$ and $\{X_i, t+1\le i \le n\}$. After utilizing the indicator function $\mbox{I}(|i-j|\ge M+1)$ to exclude the interference of $C(j-i)$ with $0<|i-j|\le M$, we estimate the two covariance structures separately from the two segments. We then compare the two covariance structures through $A_{t, M}(i, j)$, so that the expectation of $\hat{\mathcal{J}}_{n, M}$ is zero under the null hypothesis, but it is non-zero with the maximum attained at the change point under the alternative hypothesis. Finally, we choose $W_M (i, j)$ to accumulate all the structural comparisons, each of which is obtained through $A_{t, M}(i, j)$. 
%To circumvent the difficulty of selecting $t$ and $A_{t, M}(i, j)$ based on which the partitioned two samples lead to the most significant structural comparison, we use $W_M (i, j)$ to accumulate all structural comparisons. %An alternative way is to consider $\max_{M+2 \le t \le n-M-2} A_{t, M}(i, j)$. A comparison between the two approaches will be provided in Section 4.

%{\bf Remark 2.4} With $\mbox{I}(|i-j|\ge M+1)$ in $W_M (i, j)$, $X_i$ and $X_j$ are temporally independent. We thus eliminate the effect of temporal dependence $C(j-i)$ for $i \ne j$ and obtain the neat expressions for the expectation and variance of $\hat{\mathcal{J}}_{n, M}$ given in Propositions 1 and 2. 
%If data are temporally independent ($M=0$), the indicator function $\mbox{I}(|i-j|\ge M+1)$ in $W_M (i, j)$ is always one and thus not needed. Otherwise, it is imposed to eliminate the effect of temporal dependence $C(j-i)$ for $i \ne j$, so that the statistic is focused on detecting the change in the covariance matrix only. The effect of temporal dependence cannot be effectively eliminated by a bias correction term because $C(j-i)$ under the alternative varies with $i$ and $j$ as described in (C1).

%The following proposition demonstrates the expectation of (\ref{estimator}) which, under the alternative, is significantly different from the value under the null hypothesis. % is designed for detecting the change of covariance structure.

Since the main task is to detect change in the covariance structure, we 
assume without further notice that $\mu=0$ in (\ref{model}), and focus on $\hat{\mathcal{J}}_{n, M}$ to facilitate theoretical investigation.  
All the established results can be readily extended to $\hat{\mathcal{J}}^*_{n, M}$ with $\mu \ne 0$. 

\bigskip
{\bf Proposition 1.} Assume (\ref{model}) and (C1). Under the null hypothesis, $$\mbox{E}(\hat{\mathcal{J}}_{n, M})=0.$$ Under the alternative hypothesis,
\[
\mu_{\hat{\mathcal{J}}_{n, M}}\equiv \mbox{E}(\hat{\mathcal{J}}_{n, M})=\frac{1}{n^2}\sum_{i, j=1}^n W_M(i, j)\mbox{tr}(\Sigma_i \Sigma_j). %\label{mean}
\]

Since the expectation of $\hat{\mathcal{J}}_{n, M}$ under the alternative hypothesis differs from its expectation under the null hypothesis, it can be used to test heterogeneity of the covariance structure after we standardize it. This requires us to further derive the variance of the test statistic.

\bigskip
{\bf Proposition 2.} Under (\ref{model}) and (C1)-(C2), as $p \to \infty$, the  variance of $\hat{\mathcal{J}}_{n, M}$ is
\[
\sigma_{\hat{\mathcal{J}}_{n, M}}^2=\frac{4}{n^4}\sum_{i, j=1}^n \sum_{k, l=1}^n W_M(i, j) W_M(k, l)\mbox{tr}^2\{C(i-k) C(l-j)\}\{1+o(1)\}.
%\label{variance}
\]

Under the null hypothesis, (C1) assumes that the spatial and temporal dependence is stationary. The leading order variance can be simplified as
\be
\sigma_{\hat{\mathcal{J}}_{n, M}, 0}^2=\frac{4}{n^4}\sum_{i, j=1}^n \sum_{h_1, h_2}W_M(i, j) W_M(i-h_1, j+h_2)\mbox{tr}^2\{C(h_1) C(h_2)\},\label{variance_null}
\ee
where $h_1, h_2 \in \{0, \pm 1, \cdots, \pm M\}$.

\subsection{Stopping rule}

We intend to propose a stopping rule based on $\hat{\mathcal{J}}_{n, M}$ (\ref{estimator}) for the hypotheses (\ref{Hypo}). However, there are two issues we need to address when using $\hat{\mathcal{J}}_{n, M}$. The first issue is nuisance parameters. There are two nuisance parameters related with $\hat{\mathcal{J}}_{n, M}$: the $M$ for temporal dependence and the standard deviation of $\hat{\mathcal{J}}_{n, M}$ under the null hypothesis. While $M$ is zero under temporal independence and $\sigma_{\hat{\mathcal{J}}_{n, M}, 0}$ in (\ref{variance_null}) becomes $2p /n^2 \sqrt{\sum_{i,j} W_M^2(i, j)}$ without temporal dependence and with the identity covariance matrix $\Sigma$, they are unknown in the presence of spatial and temporal dependence. 
Similar to Pollak and Siegmund (1991), we consider a training sample of size $n_0$ to provide estimation of both nuisance parameters. Estimating $M$ based on the training sample is covered in Section 3.4. To estimate the standard deviation of $\hat{\mathcal{J}}_{n, M}$ under the null hypothesis, we need to estimate $\mbox{tr}\{C(h_1) C(h_2)\}$ because it is the only unknown in (\ref{variance_null}). Based on a training sample, we estimate it by   
\begin{eqnarray}
\widehat{\mbox{tr}\{C(h_1)C(h_2)\}}=\frac{1}{n^*}\sum_{s,t}^*X_{t+h_2}^T X_s X_{s+h_1}^T X_t, \label{var.est}
\end{eqnarray}
where $\sum^*$ represents the sum of indices that are at least
 $M$ apart in the training sample, and $n^*$ be the corresponding number of indices. The consistency of the estimator is established in Theorem 3 of Section 3.3.  

The second issue related with $\hat{\mathcal{J}}_{n, M}$ is the computational complexity. From (\ref{estimator}), $\hat{\mathcal{J}}_{n, M}$ involves the weight function $W_M(i, j)$ which sums $t$ from $M+2$ to $n-M-2$. As mentioned in Remark 2.2, at each $t$, the test statistic needs to compare the two covariance structures estimated separately from the two segments $\{X_i, 1\le i \le t \}$ and $\{X_i, t+1\le i \le n\}$. 
When $n$ is large, it can be time consuming to compute $\hat{\mathcal{J}}_{n, M}$. To reduce the computational time, we consider a modified statistic    
\[
\hat{\mathcal{J}}_{n, M, H}= \frac{1}{H^2} \sum_{i, j=n-H+1}^n  W_{M}(i, j)(X_i^T X_j)^2,
\]
which, compared with the original $\hat{\mathcal{J}}_{n, M}$, is only based on the past $H$ observations from the current time $n$ and thus can be computationally more efficient. It is not rare to utilize a moving window $H$ for the online change point detection; see, for example, Lai (1995), Cao et al. (2019) and Chen (2019). While the motivation is to reduce the computational complexity, the impact of imposing $H$ on the proposed method needs to be carefully addressed. We display its effect on our stopping rule explicitly through asymptotic results in next section and simulation studies in Section 4. Some other guidelines in selecting the window size can be seen in Lai (1995).  
%Its consistency will be established in Theorem 3 in Section 3.3. 

We are now ready to propose the stopping rule 
\begin{eqnarray}
T_H(a, M)= \mbox{inf} \biggl\{n-n_0:  \left | \frac{\hat{\mathcal{J}}_{n, M, H}}{\hat{\sigma}_{{n_0, M, H}}} \right | > a, \,\, n > n_0 \biggr\}, \label{stopping-rule}
\end{eqnarray}
where $n_0$ is the size of a training sample, and $\hat{\sigma}^2_{{n_0, M, H}}$ is the estimator of the variance of $\hat{\mathcal{J}}_{n, M, H}$ under the null hypothesis. Using (\ref{var.est}), we obtain  
\be
\hat{\sigma}^2_{{n_0, M, H}}=\frac{4}{H^4}\sum_{i, j=1}^H \sum_{h_1, h_2} W_M(i, j) W_M(i-h_1, j+h_2)\widehat{\mbox{tr}^2\{C(h_1)C(h_2)\}}. \label{var.est2}
\ee

The proposed stopping rule terminates the detection procedure in a minimal number of new observations after the training sample $n_0$, when the absolute value of the standardized test statistic is above the threshold $a$.  
For online change point detection, $a$ should be chosen to balance the tradeoff between false alarms and detection delay. If $a$ is too small, the stopping rule may detect a change quickly but unavoidably generate a lot of false alarms if there is no change. On the other hand, if $a$ is too large, the desire to avoid false alarms will lead to a significant delay between the change point and the termination time. A conventional method for choosing $a$ is that the average run length (ARL) can be controlled at any pre-specified value. In next section, we establish an explicit relationship between $a$ and ARL in Theorem 1, which allows us to quickly determine $a$ with no need to run time-consuming Monte Carlo simulations. The proposed stopping rule also depends on the temporal dependence $M$, which is unknown in practice. In Section 3.4, we provide an algorithm to consistently estimate $M$. Our investigation show that the stopping rule based on the estimated $M$ performs as well as that based on the true $M$.

%{\bf Remark 2.5} When there is no change, ${\hat{\mathcal{J}}_{n, M, H}}/{\hat{\sigma}_{\hat{\mathcal{J}}_{n, M, H}, 0}}$ follows an asymptotic standard normal distribution as shown in Appendix. If there is a change point $\tau \in [n_0, n]$, we would expect $|{\hat{\mathcal{J}}_{n, M, H}}/{\hat{\sigma}_{\hat{\mathcal{J}}_{n, M, H}, 0}}|$ to be large according to Proposition 1. With a proper $a$, an alarm is triggered with the minimum $n$ that makes $|{\hat{\mathcal{J}}_{n, M, H}}/{\hat{\sigma}_{\hat{\mathcal{J}}_{n, M, H}, 0}}|$ above the threshold $a$.

%The comparison of two stopping rules will be given in Section 4.

\section{Asymptotic Results}

\subsection{Average run length}

%In the stopping rule (\ref{stopping-rule}), the threshold needs to be chosen so that the expected average run length can be controlled at a pre-specified level. To determine the value of the threshold, we need to establish the asymptotic cumulative distribution function of the stopping time $T_H(a)$ by the following theorem.
 Let $\E_{\infty}$ and $\mbox{P}_{\infty}$ denote the expectation and probability, respectively, under the null hypothesis. Let
\[
g(t/H, a)={2\log (t/H)}+1/2 \log \log (t/H)+\log (4/\sqrt{\pi})-a\sqrt{2\log (t/H)}.
\]

The ARL is defined to be the expected value of the stopping time under the null hypothesis. %and $\dot{g}(t/H, a)$ denote the derivative with respect to $t$.
The following theorem establishes the ARL or $\E_{\infty}\{T_H(a, M) \}$ for the proposed stopping rule (\ref{stopping-rule}).

\medskip
{\bf Theorem 1.} Assume (\ref{model}) and (C1)-(C2). As $p \to \infty$, and both $H$ and $a$ $\to \infty$ satisfying $H=o\{\mbox{exp}(a^2/2)\}$, 
\[
\mbox{E}_{\infty}\{T_H(a, M) \}=\biggl( H+ \int_{H}^{\infty} \mbox{exp}\biggl[-2 \mbox{exp}\biggl\{g(t/H, a)\biggr\}\biggr] dt \biggr)\{1+o(1)\}.
\]

As shown in the proof of Theorem 1, the ARL is readily obtained by establishing the cumulative distribution function of $T_H(a, M)$ as $a \to \infty$. Since the randomness of $T_H(a, M)$ is determined by ${\hat{\mathcal{J}}_{n, M, H}}/{\hat{\sigma}_{{n_0, M, H}}}$, the cumulative distribution of the former can be derived by establishing the asymptotic distribution of the latter when $p$ and $H \to \infty$. Here the condition $H=o\{\mbox{exp}(a^2/2)\}$ specifies the growth rate of $H$ with respect to $a$. It is imposed to ensure that the probability the procedure stops within the window $H$ goes to zero exponentially fast.   
%\[
%\mbox{P}_{\infty}\{T_H(a, M)\le t \}=1-\mbox{exp}\biggl[-2 \mbox{exp}\biggl\{g(t/H, a)\biggr\}\biggr].
%\]

Theorem 1 states that the ARL depends on the threshold $a$ and the window-size $H$. In particular, it increases as $a$ increases. This can also be seen from the proposed stopping rule (\ref{stopping-rule}), where raising $a$ makes the standardized test statistic less likely to go beyond the $a$ when there is no change point. The practical usefulness of Theorem 1 is that with any pre-specified ARL and $H$, we can quickly determine the value of $a$ by solving the equation rather than running time-consuming Monte Carlo simulations. %The accuracy of Theorem 1 is confirmed by Monte Carlo simulations in Section 4.1.

\subsection{Expected detection delay}
When there is a change point $\tau$, the proposed stopping rule is conventionally examined by the expected detection delay (EDD), $\mbox{E}_{\tau}\{T_H(a, M)-(\tau-n_0)| T_H(a, M)>\tau-n_0\}$ with $\tau \ge n_0$. %which is the expected number of additional observations after the change point if the process is terminated after the change point. 
In the literature, it is customary to consider the EDD for the so-called immediate change point; see, for example, Siegmund and Venkatraman (1995) and Xie and Siegmund (2013). In terms of our configuration, it refers to the change occurs immediately after the training sample $n_0$ and the corresponding EDD is 
$\mbox{E}_0 \{T_H(a, M)\}$.  
The main reason to consider the EDD of the immediate change point is that for many stopping rules, the supremum of all the EDDs attains at the immediate change point. It is therefore important to see if such property is still held by our proposed stopping rule. 
We establish the following theorem which confirms this conclusion. More importantly, the theorem provides an upper bound for the EDDs.

\medskip
{\bf Theorem 2.} Assume (\ref{model}) and (C1)-(C2). Consider the change point $\tau \ge n_0$. As $p \to \infty$, and both $H$ and $a \to \infty$ satisfying $a=O(H^r)$ with ${1}/{2} \le r <1$, 
\[
\mbox{sup} \,\,\mbox{E}_{\tau}\{T_H(a, M)-(\tau-n_0)| T_H(a, M)>\tau-n_0\}=\mbox{E}_0\{T_H(a, M)\}, \quad \mbox{and}
\]
\[ 
\mbox{E}_0\{T_H(a, M)\}\leq (M+2)+ \frac{\sqrt{a \cdot H \cdot \sigma_{H, M, 0}}}{||\Sigma_{\tau+1}-\Sigma_{\tau}||_F}
\{1+o(1)\},
\]
where $\sigma_{H, M, 0}$ is obtained by replacing $n$ with $H$ in (\ref{variance_null}), and $||\cdot ||_F$ represents the matrix Frobenius Norm.  

\medskip
%Both the upper bound of the stopping time $T_H(a, M)$ and the threshold $a$ are allowed to diverge with the window length $H$ at the rates $o(H)$ and $O(H^r)$ for ${1}/{2} \le r <1$, respectively. In particular, the condition $a=O(H^r)$ is equivalent to requiring that the ARL diverges with $H$ at a relatively fast rate, the value of which can be obtained by solving the equation in Theorem 1.  
Theorem 2 demonstrates the impact of some key factors on the EDD. First, a larger $M$ could lead to a greater EDD, showing the adverse effect of the dependence on change-point detection. Second, the impact of the threshold $a$ on the EDD essentially depends on the choice of ARL,  because $a$ is obtained by solving the equation in Theorem 1 in which 
the window size $H$ and ARL are pre-specified by the user. Generally speaking, a larger user-chosen ARL leads to a higher value of $a$ and thus a greater EDD. Finally, the impact of $\sigma_{H, M, 0}^{1/2} \,||\Sigma_{\tau+1}-\Sigma_{\tau}||_F^{-1}$ can be demonstrated by applying the result $\sigma_{H, M, 0}^{1/2}=O(||\Sigma_{\tau}||_F)$ from the proof of Theorem 1 to obtain $$\frac{\sqrt{\sigma_{H, M, 0}}}{||\Sigma_{\tau+1}-\Sigma_{\tau}||_F}=O\biggl( \frac{||\Sigma_{\tau}||_F}{||\Sigma_{\tau+1}-\Sigma_{\tau}||_F}\biggr).$$ The result shows that the EDD can be significantly reduced by increasing the ratio of the change in covariance structure to the original covariance.       

\smallskip
{\bf Remark 3.1} It requires a minimum change in the covariance structure, for the proposed stopping rule to detect the change point. To understand this, we consider the configuration with the immediate change after the training sample. As the window continuously moves to the right, the number of observations with $\Sigma_{\tau}$ decreases but the number of observations with $\Sigma_{\tau+1}$ increases.    
If the detection procedure has not yet stopped when the last observation with $\Sigma_{\tau}$ begins to leave the window, it probably won't be able to stop because the process ends up with all the $H$ observations having the same $\Sigma_{\tau+1}$. Theorem 2 actually provides a minimum change the proposed stopping rule requires. By noticing that the right hand side of the inequality in Theorem 2 must be no more than $H$, the change in covariance structure  
 \[
||\Sigma_{\tau+1}-\Sigma_{\tau}||_F \ge \sqrt{{a}/{H}} ||\Sigma_{\tau}||_F, 
\] 
where $\sqrt{a/H}||\Sigma_{\tau}||_F$ is therefore the minimum change in the covariance structure the proposed stopping rule is able to detect. To provide an insight of the result, we consider $\Sigma_{\tau}=I_p$ where $p=1000$, and $\Sigma_{\tau+1}=(\rho^{|i-j|})$ where $0 < \rho <1$. Further, we choose $H=100$ and obtain $a=3.58$ by solving the equation in Theorem 1 so that the ARL is controlled around $5000$.  Solving $||\Sigma_{\tau+1}-\Sigma_{\tau}||_F = \sqrt{{a}/{H}} ||\Sigma_{\tau}||_F$, we obtain the minimum $\rho$ for the stopping rule to detect the change is  $0.133$.    

\smallskip
{\bf Remark 3.2} The proposed stopping rule is based on the $L_2$-norm  statistic. When $\Sigma_{\tau+1}$ differs from $\Sigma_{\tau}$ in a large number of components, the stopping rule is advantageous as the detection delay can be significantly reduced by accumulating all the differences through $||\Sigma_{\tau+1}-\Sigma_{\tau}||_F$. When $\Sigma_{\tau+1}$ differs from $\Sigma_{\tau}$ only in a sparse number of components, the components without the change do not contribute to $||\Sigma_{\tau+1}-\Sigma_{\tau}||_F$ but to $||\Sigma_{\tau}||_F$, which may lead to a large $||\Sigma_{\tau}||_F/||\Sigma_{\tau+1}-\Sigma_{\tau}||_F$ and thus a long detection delay. To reduce the detection delay under the sparse situation, we can rewrite the test statistic in the stopping rule into
$$\hat{\mathcal{J}}_{n, M, H}= \frac{2}{H^2}\sum_{1\le k \le l \le p} \sum_{i, j=n-H+1}^n  W_{M}(i, j)Y_{i, kl} Y_{j, kl},$$
where $Y_{i, kl}=X_{i, k} X_{i, l}$ and $X_{i, k}$ is the $k$th component of the $p$-dimensional random vector $X_i$, and remove the elements $Y_{i, kl} Y_{j, kl}$ with no change. Since the screening must be conducted through a data-driven approach, its effect on the ARL and EDD deserves some future research efforts. %An alternative way is to apply the idea in Xie and Siegmund (2013) to assign a fraction $p_0$ of the components with the change. As suggested in Xie and Siegmund (2013), the sensitivity of the stopping rule to the assigned $p_0$ needs to be addressed.               

%Theorem 2 states that if the magnitude of change in the covariance structure is strong, the stopping rule can choose a short window size $H$ to detect the change. Otherwise, it needs a longer window size $H$. To demonstrate the idea, we simply consider $\Sigma_{\tau}$ to be the identity matrix $I_p$. Let $\Sigma_{\tau+1}-\Sigma_{\tau}=(\delta_{ij})$ where $\delta_{ij}$ is the difference between $\Sigma_{\tau+1}$ and $\Sigma_{\tau}$ at $i$th row and $j$th column. Let the number of nonzero $\delta_{ij}$ be $[g^{1-\beta} ]$ where $[\cdot ]$ is the integer truncation function, $g=p(p+1)/2$ and $\beta$ is the sparsity parameter taking the value on $[0, 1]$.     
%The performance of the above finite-dimension corrected theoretical upper bound for EDD is also presented in numerical studies.

\subsection{Training sample}

%To implement the stopping rule, we need a training sample which has no change point in covariance structure. 
A training sample primarily provides estimation of unknown nuisance parameters for the proposed stopping rule. Because of its importance, it is worth discussing the availability of the training sample in practice. In many biological studies, prior regulatory networks or pathway information for different biological processes are available through massive datasets (Li and Li, 2008). Such datasets can be used as a training sample  if the contained prior knowledge or information matches the initial covariance structure of the considered online detection process. Under other circumstances, a training sample can be historical observations from previous experimental runs subject to similar experimental conditions, after their stationarity of the covariance structure has been confirmed. Suppose $\{X_i, 1\le i \le n_0\}$ are such historical observations. To check their stationarity in the covariance structure, it is equivalent to considering the hypotheses   
\begin{eqnarray}
&H^*_0&: \Sigma_1=\cdots=\Sigma_{n_0}, \qquad \mbox{against} \nonumber\\
&H_1^*&: \Sigma_1=\cdots=\Sigma_{\tau_1}\ne \Sigma_{\tau_1+1}=\cdots=\Sigma_{\tau_q}\ne \Sigma_{\tau_q+1}=\cdots=\Sigma_{n_0}, \label{Hypo2}
\end{eqnarray}
where $1 \le \tau_1 < \cdots < \tau_q <n_0$ are unknown change points. %This is an offline testing problem as the sample has been collected. %If we fail to reject the null hypothesis, $\{X_i, 1\le i \le n_0\}$ serves as a training sample. 
To test the null hypothesis, we consider the test statistic $\hat{\mathcal{J}}_{n_0, M}$ which is obtained by replacing $n$ with $n_0$ in (\ref{estimator}). The rationale of using $\hat{\mathcal{J}}_{n_0, M}$ is that its expectation can distinguish the alternative from the null hypothesis. The following theorem establishes the asymptotic normality of $\hat{\mathcal{J}}_{n_0, M}$.

\medskip
{\bf Theorem 3.} Assume (\ref{model}) and (C1)-(C2). As $n_0 \to \infty$,
\[
\frac{\hat{\mathcal{J}}_{n_0, M}-\mu_{\hat{\mathcal{J}}_{n_0, M}}}{\sigma_{\hat{\mathcal{J}}_{n_0, M}}} \xrightarrow{d} N(0, 1),
\]
where $\mu_{\hat{\mathcal{J}}_{n_0, M}}$ and $\sigma_{\hat{\mathcal{J}}_{n_0, M}}$ are given by Propositions 1 and 2, respectively, with $n$ replaced by $n_0$. In particular, under $H_0$ of (\ref{Hypo2}),
\[
\frac{\hat{\mathcal{J}}_{n_0, M}}{\hat{\sigma}_{\hat{\mathcal{J}}_{n_0, M},0}} \xrightarrow{d} N(0, 1),
\]
where $\hat{\sigma}_{\hat{\mathcal{J}}_{n_0, M},0}$ is defined in (\ref{var.est2}) with $n$ replaced by $n_0$.

From Theorem 3, we reject $H^*_0$ of (\ref{Hypo2}) with a nominal significance level $\alpha$ if $\hat{\mathcal{J}}_{n_0, M}/\hat{\sigma}_{\hat{\mathcal{J}}_{n_0, M},0}> z_{\alpha}$, where $z_{\alpha}$ is the upper $\alpha$-quantile of the standard normal. Otherwise, we fail to reject $H^*_0$ and hereby obtain a training sample for the proposed stopping rule. %It thus provides a practical way of testing whether there exists any change of the covariance structure in the training sample.

\subsection{Stopping rule with estimated $M$}

The unknown $M$ in the stopping rule (\ref{stopping-rule}) can be estimated through the training sample $X_1, \cdots, X_{n_0}$. From (C1), we know that $\mbox{Cov}(X_i, X_j)=C(i-j)$ is zero if and only if $|i-j| > M$, or equivalently, $\mbox{tr}\{C(h)C^T(h)\}$ is zero if and only if $|h| > M$. We thus  estimate $M$ through the following steps. 
\begin{itemize}
\item Using (\ref{var.est}), we compute $\widehat{\mbox{tr}\{C(h)C(-h)\}}/\widehat{\mbox{tr}\{C(0)C(0)\}}$ with $h$ starting from $0$.  
\item We terminate the process when the first non-negative integer $h^*$ satisfies
\[
\frac{\widehat{\mbox{tr}\{C(h^*)C(-h^*)\}}}{\widehat{\mbox{tr}\{C(0)C(0)\}}}\le \epsilon,
\]
where $\epsilon$ is a small constant and can be chosen to be $0.05$ in practice.  
\item We then estimate $M$ by $\hat{M}=h^*-1$. 
\end{itemize}

Let $T_H(a, \hat{M})$ be the stopping rule obtained by replacing $M$ with $\hat{M}$ in (\ref{stopping-rule}). The following theorem shows that $T_H(a, \hat{M})$ performs as well as $T_H(a, {M})$ under both null and alternative hypotheses.  

\medskip
{\bf Theorem 4.} Assume the same conditions in Theorems 1 and 2. As the training sample size $n_0 \to \infty$,
\[
\E_{\infty}\{T_H(a, \hat{M}) \}=\E_{\infty}\{T_H(a, M) \}, \,\,\,\,\,\E_{0}\{T_H(a, \hat{M}) \}=\E_{0}\{T_H(a, M) \}.
\]
%where the expression of $\E_{\infty}\{T_H(a, M) \}$ is specified in Theorem 1, and the upper bound of $\E_{0}\{T_H(a, M) \}$ is established in Theorem 2. 

%According to (C1), $M$ separates dominant temporal dependence from the remainder.  As demonstrated in the simulation studies of Section 4 and the supplementary material, if data are dependent ($M \ne 0$), wrongly applying the procedure based on the assumption that $M=0$ can cause severe type I error and thus produce a lot of false positives when estimating locations of change points. On the other hand, choosing a value that is larger than the actual $M$  will reduce the power of the test and thus generate many false negatives. %Choosing a proper $M$ is therefore crucial for the proposed method.

\section{Simulation Studies}

\subsection{Accuracy of the theoretical ARL}

\begin{table}[t!]
\tabcolsep 5.5pt
\begin{center}
\caption{The comparison between theoretical ARLs and Monte Carlo ARLs based on $1000$ simulations. For each ARL and window-size $H$, the threshold $a$ is obtained by solving the equation in Theorem 1.}
\label{case1}
\begin{tabular}{cccccccccccc}
\hline \multicolumn{12}{c}{$H=100$}\\
\hline Theoretical &\multicolumn{3}{c}{$p=200$}&&\multicolumn{3}{c}{$p=400$} &&\multicolumn{3}{c}{$p=1000$}\\[1mm]
\cline{2-4} \cline{6-8} \cline{10-12}  (a, ARL) & $M=0$ & $1$ &$2$ && $M=0$ & $1$ &$2$ &  & $M=0$ & $1$ & $2$\\\hline
$(3.04, 1002)$     & $1178$ & $1151$ & $1194$& & $1245$ & $1284$ & $1317$& & $1302$ & $1295$&$1335$\\
$(3.42, 3008)$     & $3067$ & $3148$ & $2986$& & $3690$ & $3614$ & $3529$& & $3850$ & $3954$&$3617$\\
$(3.58, 5038)$     & $5118$ & $4527$ & $4253$& & $5799$ & $5923$ & $5212$& & $6570$ & $6102$&$5878$\\
\hline \multicolumn{12}{c}{$H=150$}\\
\hline  &\multicolumn{3}{c}{$p=200$}&&\multicolumn{3}{c}{$p=400$} &&\multicolumn{3}{c}{$p=1000$}\\[1mm]
\cline{2-4} \cline{6-8} \cline{10-12} & $M=0$ & $1$ &$2$ && $M=0$ & $1$ &$2$ &  & $M=0$ & $1$ & $2$\\\hline
$(2.88, 1005)$      & $1044$ & $1127$ & $1149$&& $1069$ & $1173$ & $1308$& & $1145$ & $1198$&$1270$\\
$(3.29, 3033)$      & $3240$ & $3156$ & $3202$&& $3505$ & $3795$ & $3628$& & $3652$ & $3759$&$3931$\\
$(3.46, 5118)$      & $5120$ & $5097$ & $5280$&& $6083$ & $5820$ & $6156$& & $6162$ & $6586$&$6794$\\
\hline
\end{tabular}
\end{center}
\end{table}

We first evaluate the performance of the stopping rule under the null hypothesis. 
The random vectors $X_{i}$ for $i=1, 2, \cdots$ are generated from
\be
X_{i}=\sum_{l=0}^{M} \Gamma_{l}\,\epsilon_{i-l}, \label{data_g}
\ee
where the $p \times p$ matrix $\Gamma_{l}=\{0.6^{|i-j|}(M-l+1)^{-1}\}$ for $i, j=1, \cdots, p$, and $l=0,\cdots, M$. Each $\epsilon_{i}$ is a $p$-variate random vector with mean $0$ and identity covariance $I_p$, and all $\epsilon_{i}$s are mutually independent. If $M=0$, all $X_i$s are mutually independent from (\ref{data_g}) and each individual $X_i$ has the covariance matrix $\Gamma_0 \Gamma_0^T$. If $M \ne 0$, $\mbox{Cov}(X_i, X_j)= \sum_{l=0}^{M-(j-i)} \Gamma_{j-i} \Gamma_l^T$ for $j-i=0, \cdots, M$.
Here we consider the normally distributed $\epsilon_{i}$. Non-Gaussian $\epsilon_{i}$ is also considered and the obtained results are included in the supplementary material of the paper. We choose the dimension $p=200$, $400$ and $1000$, the size of historical data $n_0=200$, the window-size $H=100$ and $150$, and dependence $M=0, 1, 2$, respectively.

To examine the accuracy of the theoretical ARL, we first specify its value and obtain the corresponding $a$ by solving the equation in Theorem 1. Based on the $a$, we obtain the Monte Carlo ARL by taking the average of the stopping times from $1000$ simulations.    
Table \ref{case1} compares the theoretical ARLs with the corresponding Monte Carlo ARLs under different combinations of $H$, $p$ and $M$. All the Monte Carlo ARLs are reasonably close to the theoretical ARLs, subject to some random variations from simulations under different $M$ and $p$.  %confirming the accuracy of Theorem 1. 

\subsection{Accuracy of the upper bound for EDD}

\begin{table}[t!]
\tabcolsep 3pt
\begin{center}
\caption{The comparison between theoretical upper bounds for EDDs and Monte Carlo EDDs based on $1000$ simulations with the ARL controlled around $5000$. }
\label{case2}
%\label{tab:my-table}
\resizebox{\textwidth}{!}{%
\begin{tabular}{ccccccccccccc}
\hline

       &     & \multicolumn{3}{c}{$\rho=0.6$} &  & \multicolumn{3}{c}{$0.7$} &  & \multicolumn{3}{c}{$0.8$} \\ \cline{3-5} \cline{7-9} \cline{11-13}
          &  & $M=0$   & $1$     & $2$     &  & $M=0$   & $1$     & $2$     &  & $M=0$    & $1$     & $2$     \\ \hline
\multicolumn{13}{c}{Model (a)}                                                                                 \\ \hline          
$H=100$&Monte Carlo & 16.18   & 20.14   & 24.04   &  & 11.31   & 14.34   & 16.98   &  & 8.11    & 10.31   & 12.44   \\
&Theoretical & 20.59   & 23.63   & 25.99   &  & 16.23   & 18.79   & 20.83   &  & 12.46    & 14.61   & 16.38   \\
 \hline

$H=150$&Monte Carlo & 17.49   & 21.62   & 25.45   &  & 12.34   & 15.38   & 18.56   &  & 8.90    & 11.32   & 13.37   \\
&Theoretical & 24.36   & 28.10   & 31.04   &  & 19.11   & 22.21   & 24.70   &  & 14.59    & 17.13   & 19.22   \\
\hline
\multicolumn{13}{c}{Model (b)}                                                                                 \\ \hline
$H=100$&Monte Carlo & 4.36    & 5.84    & 7.16    &  & 3.58    & 4.71    & 5.87    &  & 3.10     & 4.13    & 5.06    \\
&Theoretical & 7.42    & 9.09    & 10.58    &  & 6.07    & 7.40    & 8.76    &  & 5.11     & 6.42    & 7.60    \\
 \hline

$H=150$&Monte Carlo & 4.79    & 6.38    & 7.68    &  & 3.85    & 5.13    & 6.58    &  & 3.27     & 4.50    & 5.32    \\
&Theoretical & 8.53    & 10.38    & 11.92    &  & 6.80    & 8.45    & 9.88    &  & 5.74     & 7.19    & 8.45    \\
\hline
\multicolumn{13}{c}{Model (c)}                                                                                 \\ \hline
$H=100$&Monte Carlo & 2.84     & 3.90    & 4.94   &  & 2.68     & 3.68    & 4.78   &  & 2.63     & 3.69    & 4.72    \\
&Theoretical & 3.04     & 4.15    & 6.23   &  & 2.89     & 3.99    & 5.05   &  & 2.78     & 3.87    & 4.92    \\
 \hline

$H=150$&Monte Carlo & 2.96     & 3.94    & 5.09   &  & 2.89     & 3.93    & 4.91   &  & 2.72     & 3.76    & 4.84    \\
&Theoretical & 3.25     & 4.40    & 5.51   &  & 3.07     & 4.20    & 5.30   &  & 2.94     & 4.05    & 5.13    \\
\hline
\end{tabular}%
}
\end{center}
\end{table}

We next evaluate the performance of the stopping rule under the alternative hypothesis. In particular, 
we examine the accuracy of the upper bound for the EDD in Theorem 2. In the simulation studies, we consider an immediate change, namely the change at $\tau=201$ immediately after the historical data of size $n_0=200$. Before the change point $\tau$, the observations $X_i$ for $i=1, \cdots, 200$ are generated from (\ref{data_g}) where $\Gamma_{l}=I(M-l+1)^{-1}$ with $I$ being the identity matrix. After the change, $\Gamma_{l}=Q(M-l+1)^{-1}$ in (\ref{data_g}) where the $p \times p$ matrix $Q$ is modeled by one of the following patterns.
\begin{description}
\item[(a).]  $Q$ satisfies $QQ^T=\Sigma$, where $\Sigma_{ij}=\rho^{|i-j|}$ for $1\le i, j\le p$.
\item[(b).]  Each row of $Q$ has only three non-zero elements that are randomly chosen from $\{1, \cdots, p\}$ with magnitude $\rho$ multiplied by a random sign.
\item[(c).]  $Q$ satisfies $QQ^T=\Sigma$, where $\Sigma_{ii}=1$ for $i=1,\cdots,p$, and $\Sigma_{ij}=\rho$ for $i \ne j$.
\end{description}

Models (a)--(c) specify the bandable, sparse and strong covariance matrices, respectively. We choose $\rho=0.6, 0.7, 0.8$ to obtain different magnitudes in the covariance change, and choose the dimension $p=1000$, the window-size $H=100$ and $150$, and dependence $M=0, 1, 2$, respectively. Moreover, the threshold $a=3.58$ when $H=100$ and $a=3.46$ when $H=150$ so that the theoretical ARL is controlled around $5000$. Table \ref{case2} compares the theoretical bound for the EDD in Theorem 2 with the corresponding Monte Carlo EDD based on $1000$ simulations. As we can see, each Monte Carlo EDD is no more than its theoretical upper bound. Furthermore, both Monte Carlo EDDs and theoretical bounds decrease as $\rho$ increases with the same $M$ and $H$, but increase as $M$ increases with the same $\rho$ and $H$. The simulation results are consistent with the theoretical findings in Theorem 2.

\begin{figure}[t!]
\begin{center}
\includegraphics[width=0.45\textwidth,height=0.45\textwidth]{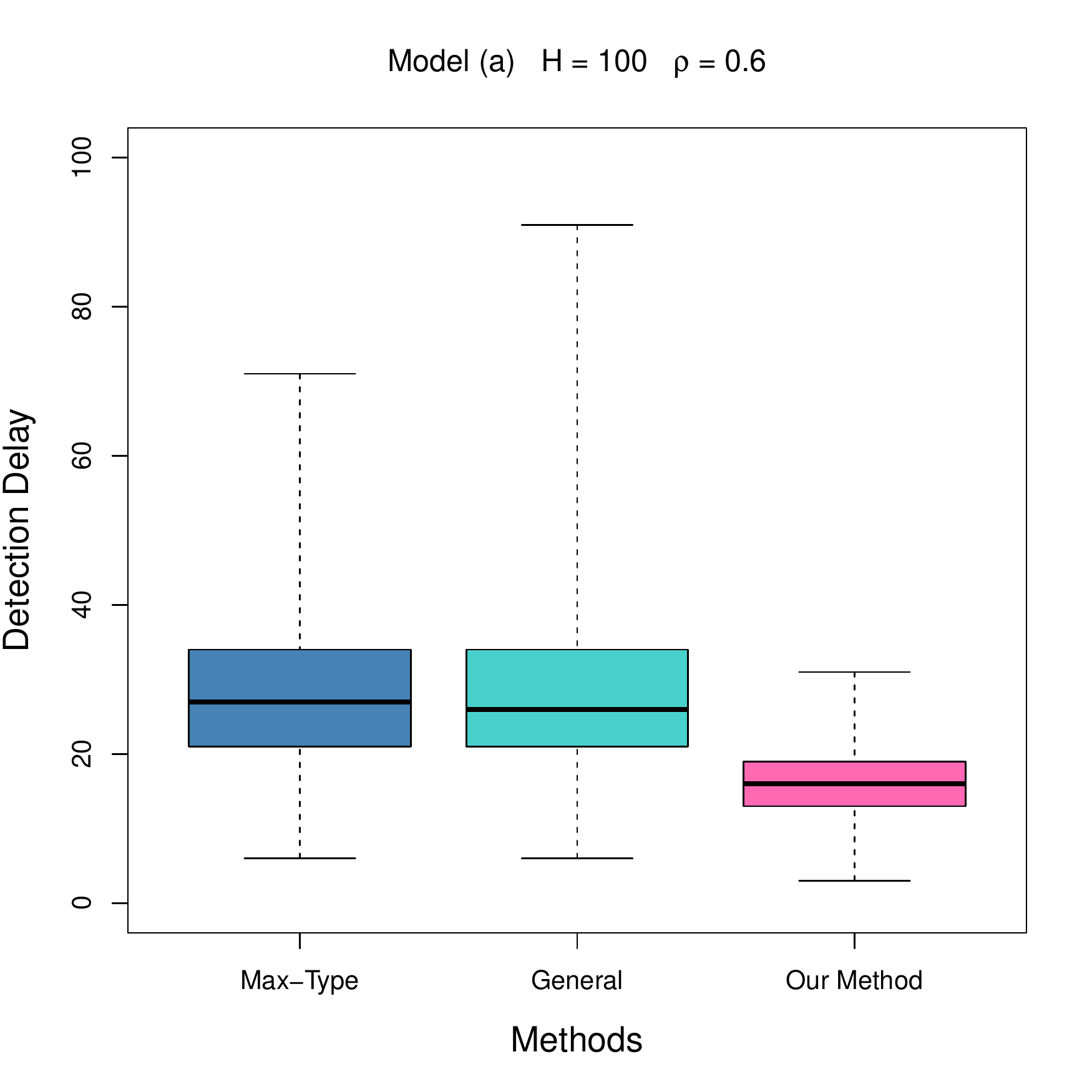}
\includegraphics[width=0.45\textwidth,height=0.45\textwidth]{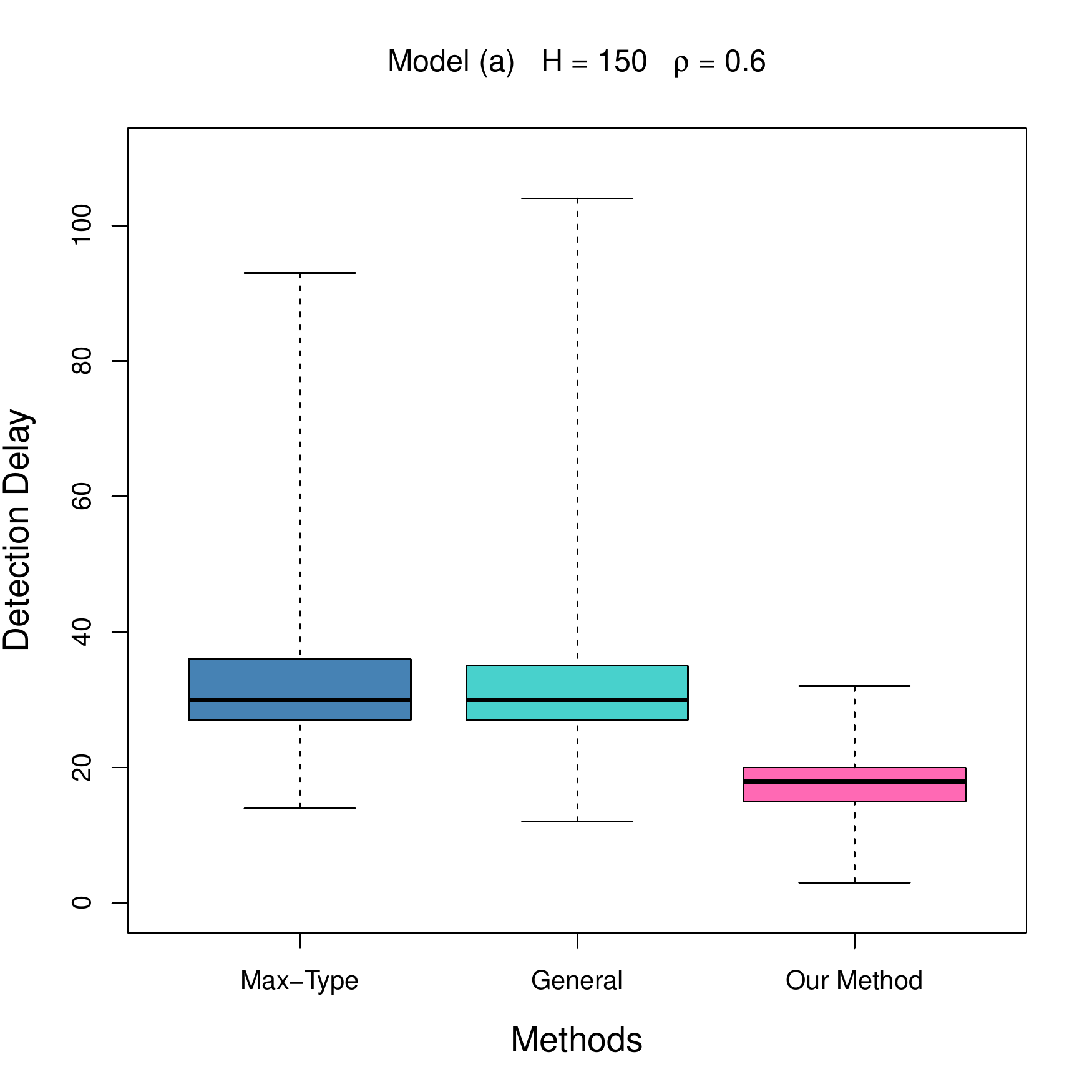}\\
\includegraphics[width=0.45\textwidth,height=0.45\textwidth]{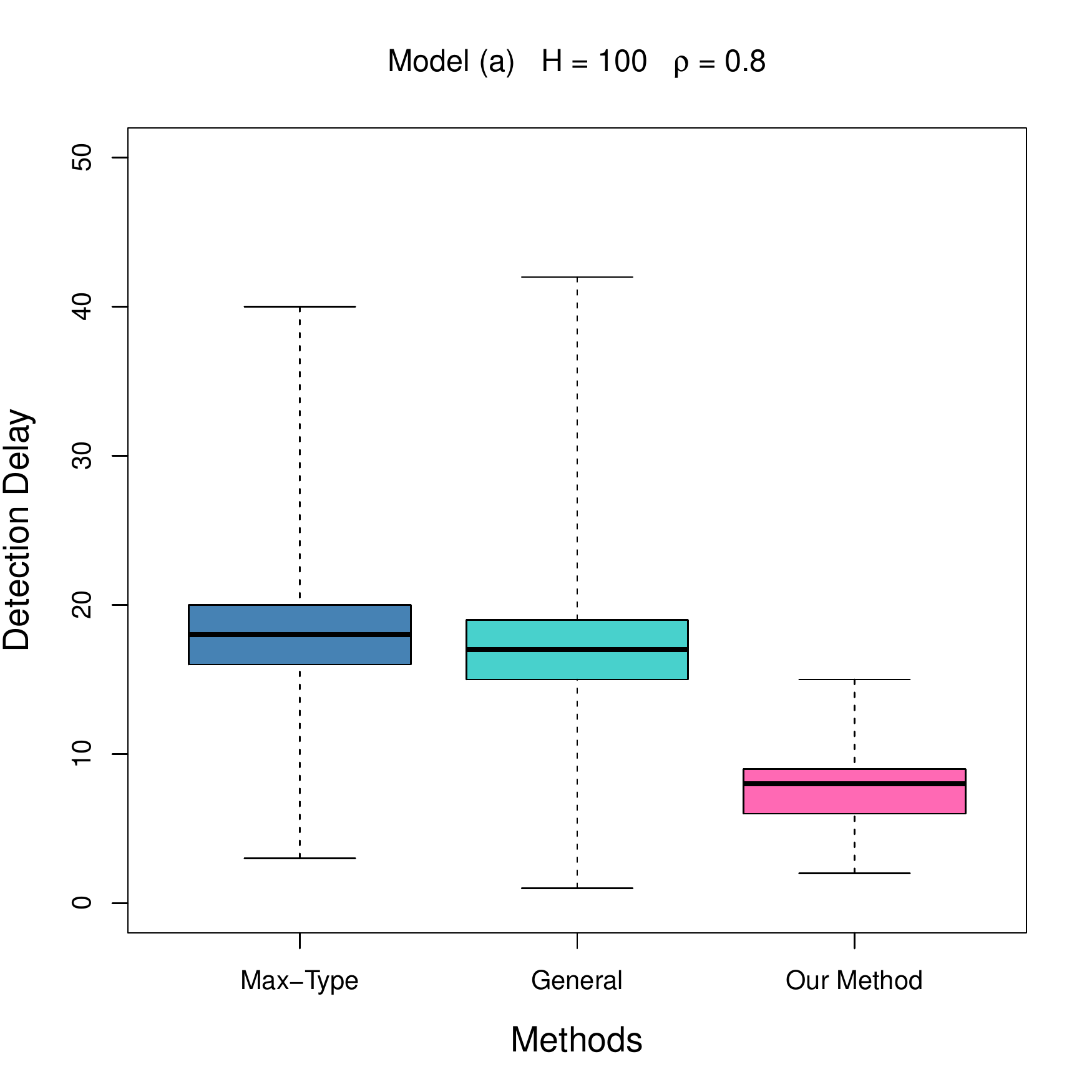}
\includegraphics[width=0.45\textwidth,height=0.45\textwidth]{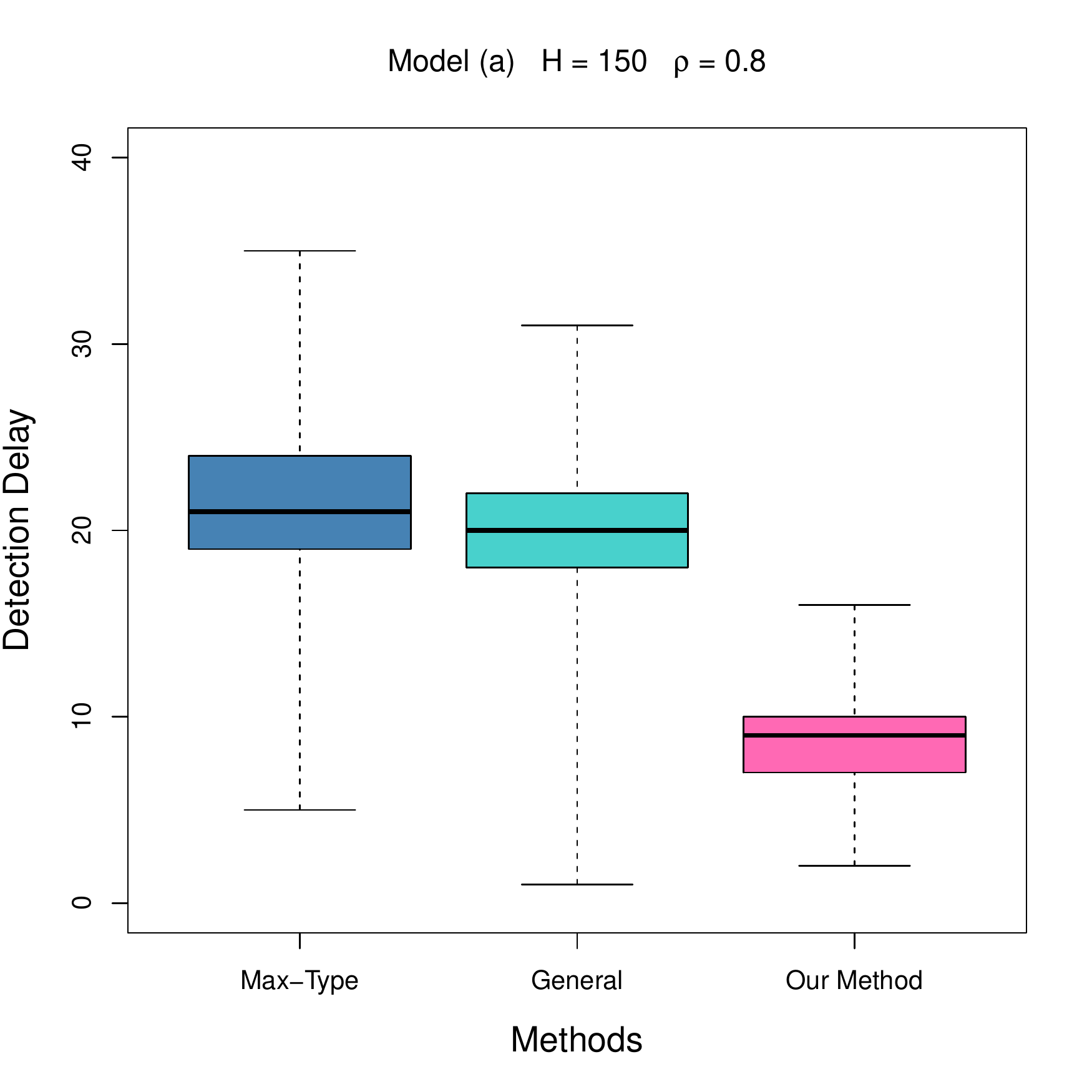}
\caption{Boxplots of detection delays for ``Max-type" and ``General" stopping rules in Chen (2019) and Chu and Chen (2018), and the proposed stopping rule. The results are based on $1000$ simulations under model (a). }
\label{comp1}
\end{center}
\end{figure}

\begin{figure}[t!]
\begin{center}
\includegraphics[width=0.45\textwidth,height=0.45\textwidth]{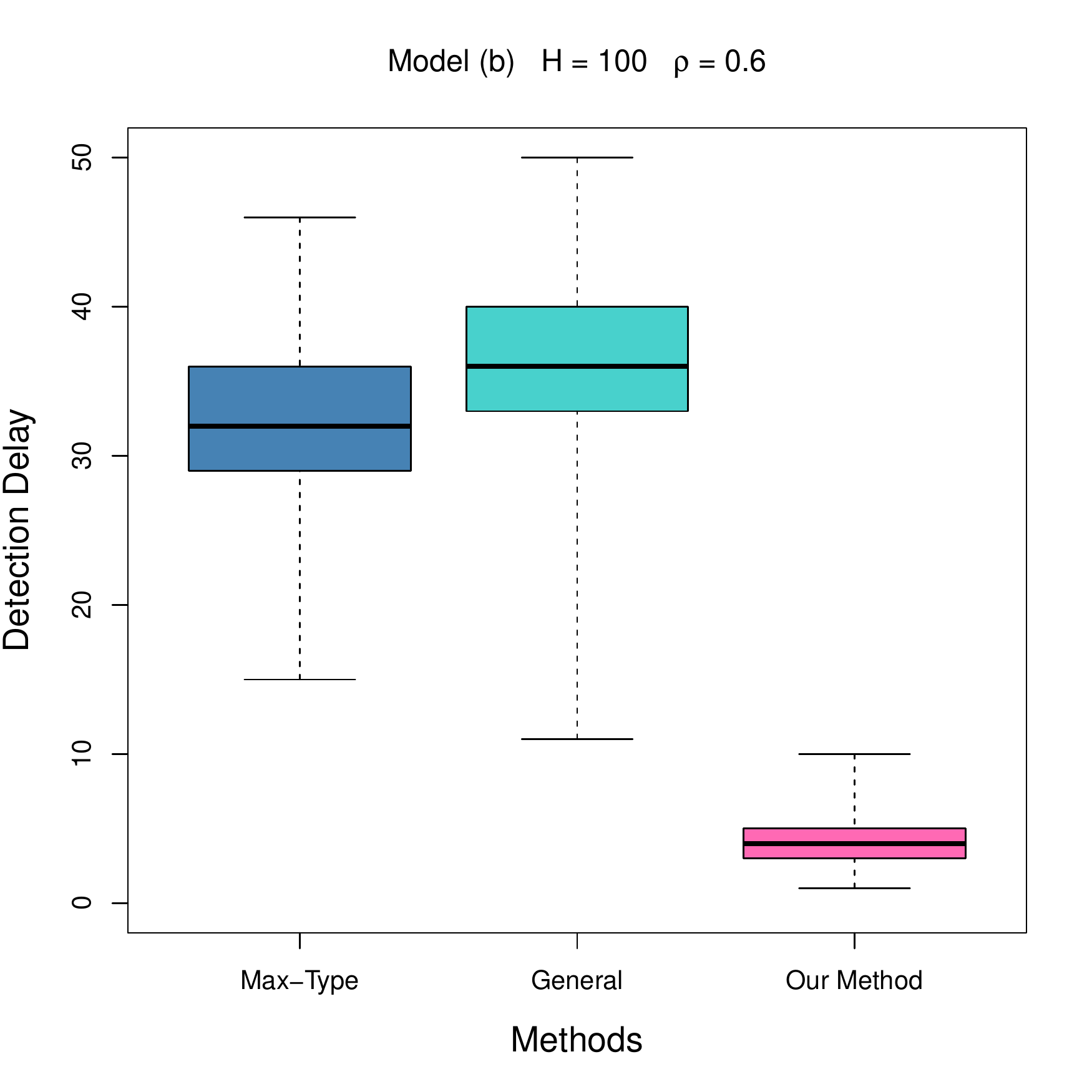}
\includegraphics[width=0.45\textwidth,height=0.45\textwidth]{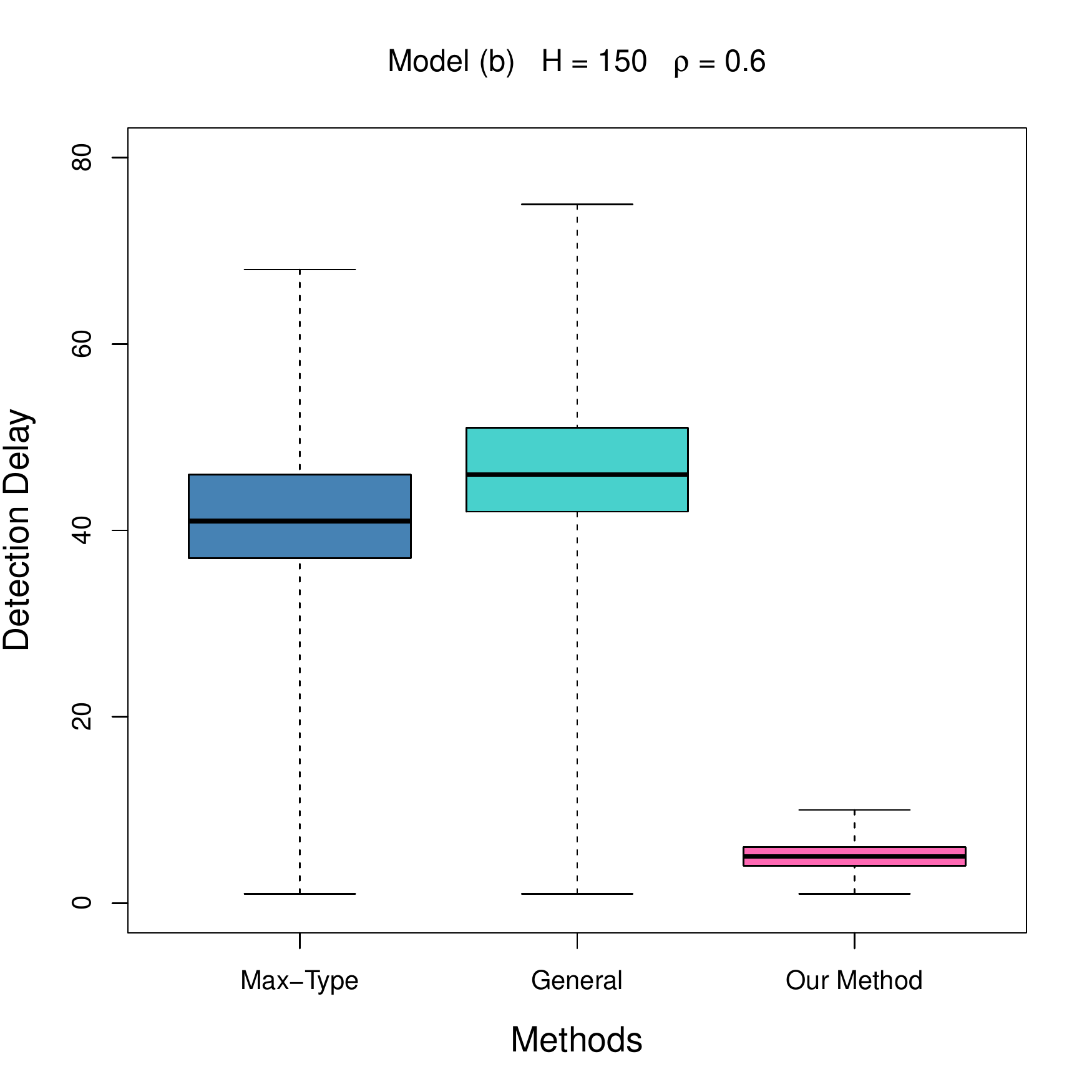}\\
\includegraphics[width=0.45\textwidth,height=0.45\textwidth]{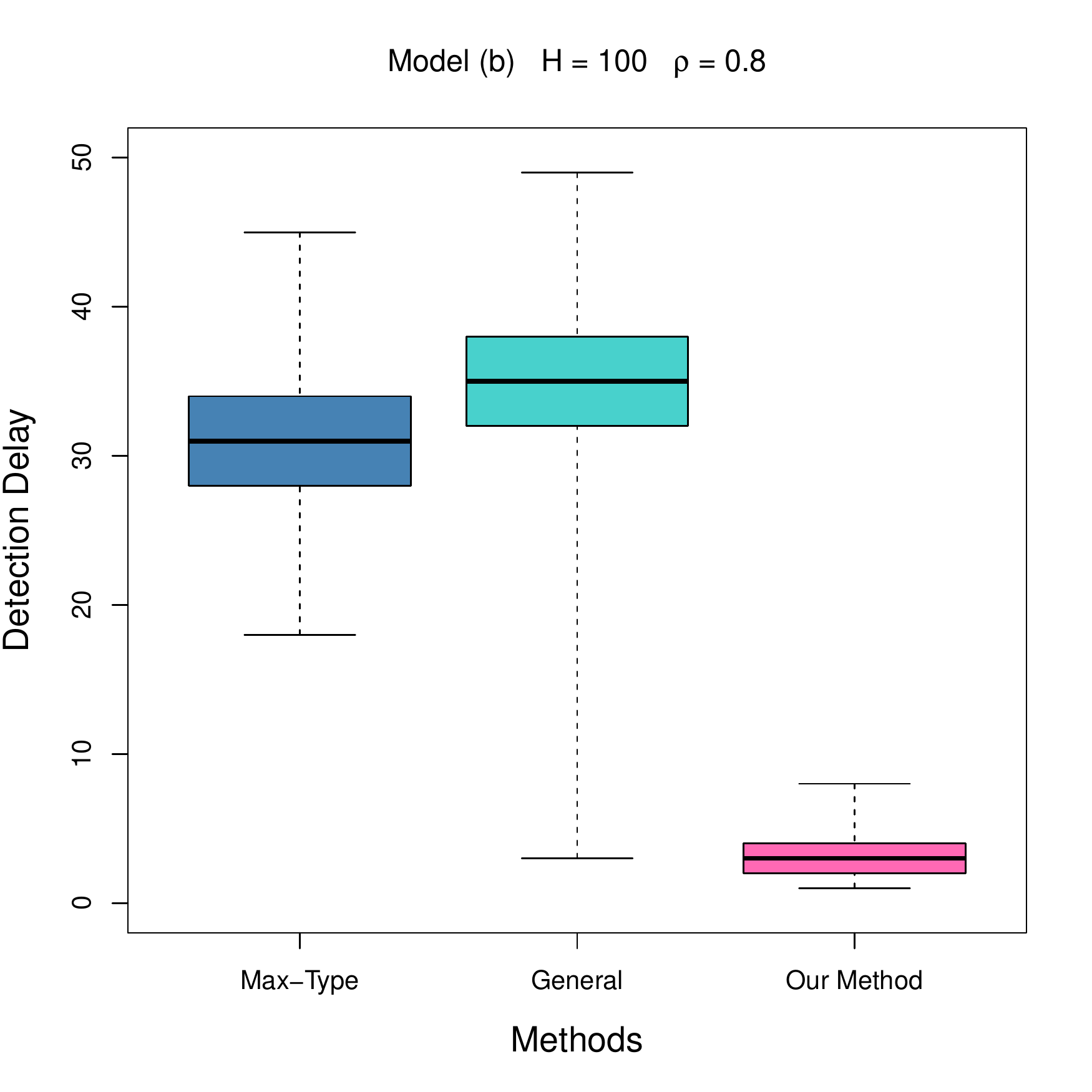}
\includegraphics[width=0.45\textwidth,height=0.45\textwidth]{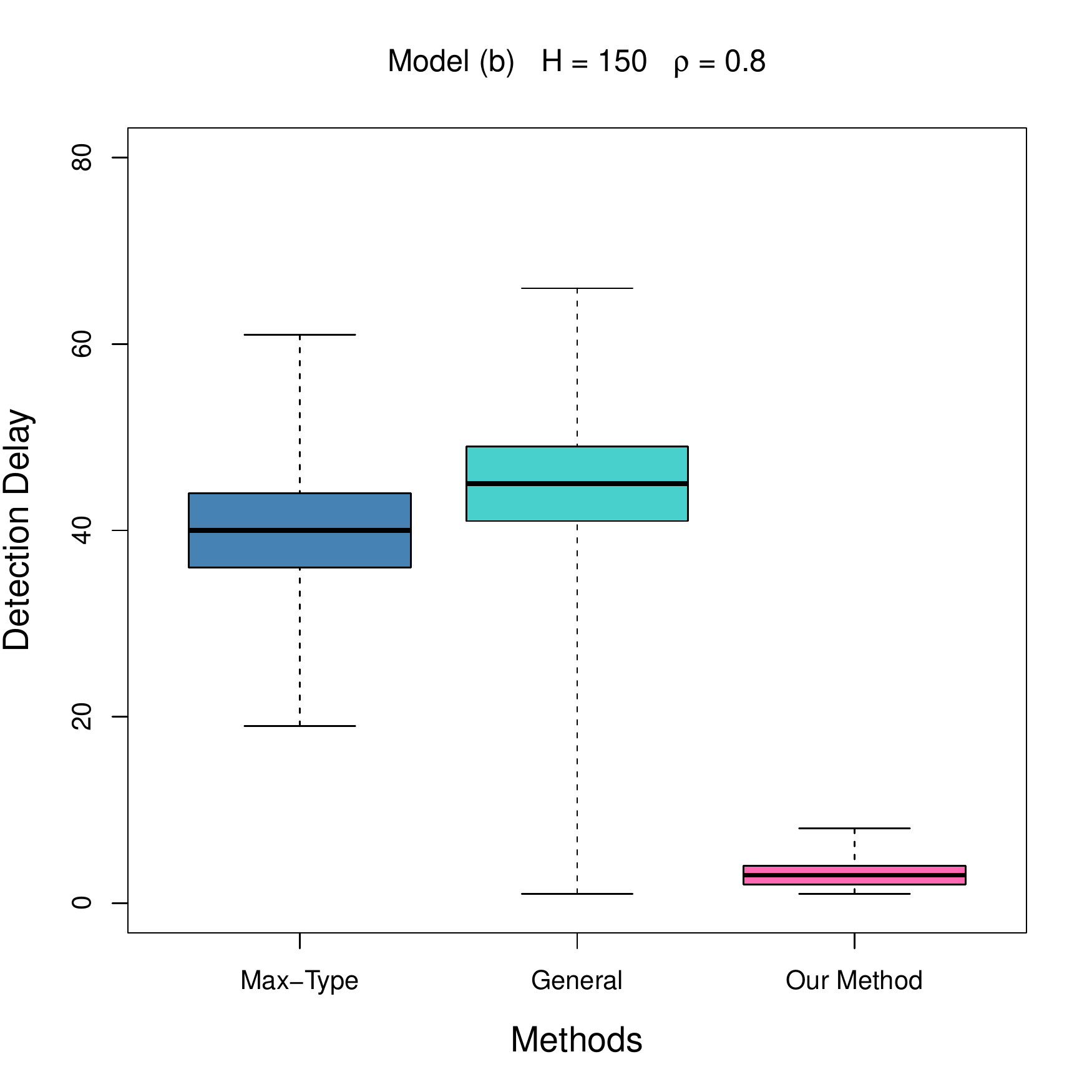}
\caption{Boxplots of EDDs for ``Max-type" and ``General" stopping rules in Chen (2019) and Chu and Chen (2018), and the proposed stopping rule. The results are based on $1000$ simulations under model (b).}
\label{comp2}
\end{center}
\end{figure}

\begin{figure}[t!]
\begin{center}
\includegraphics[width=0.45\textwidth,height=0.45\textwidth]{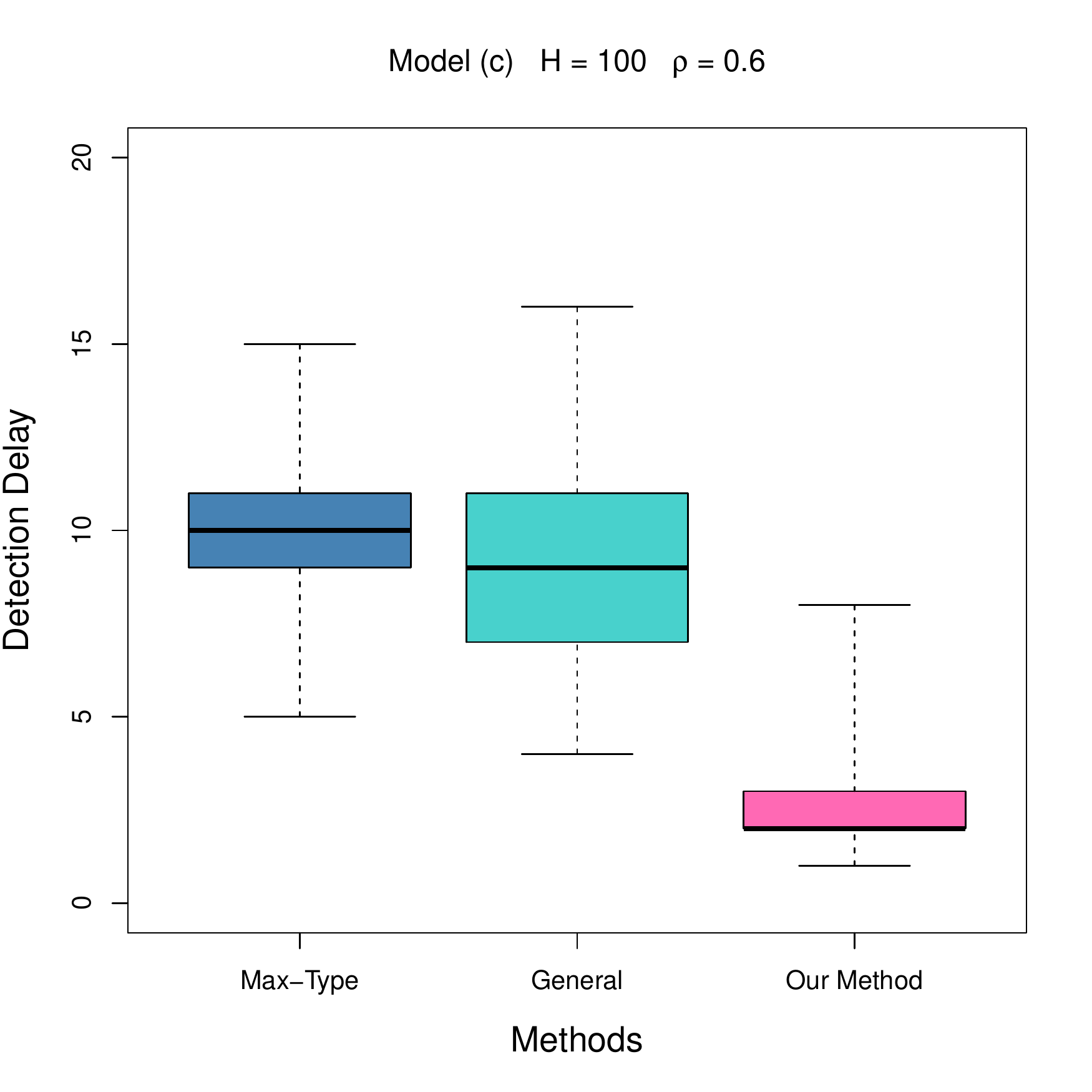}
\includegraphics[width=0.45\textwidth,height=0.45\textwidth]{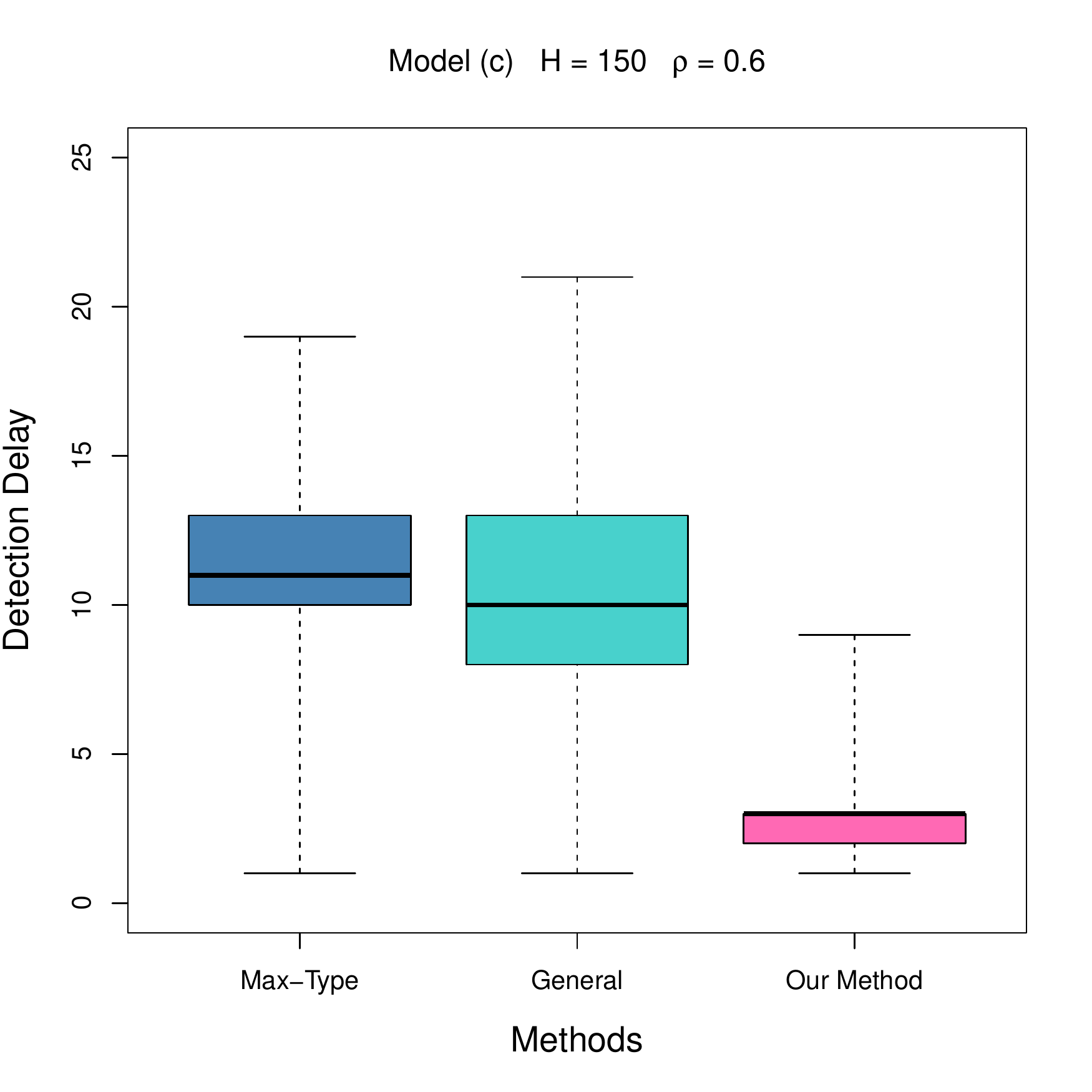}\\
\includegraphics[width=0.45\textwidth,height=0.45\textwidth]{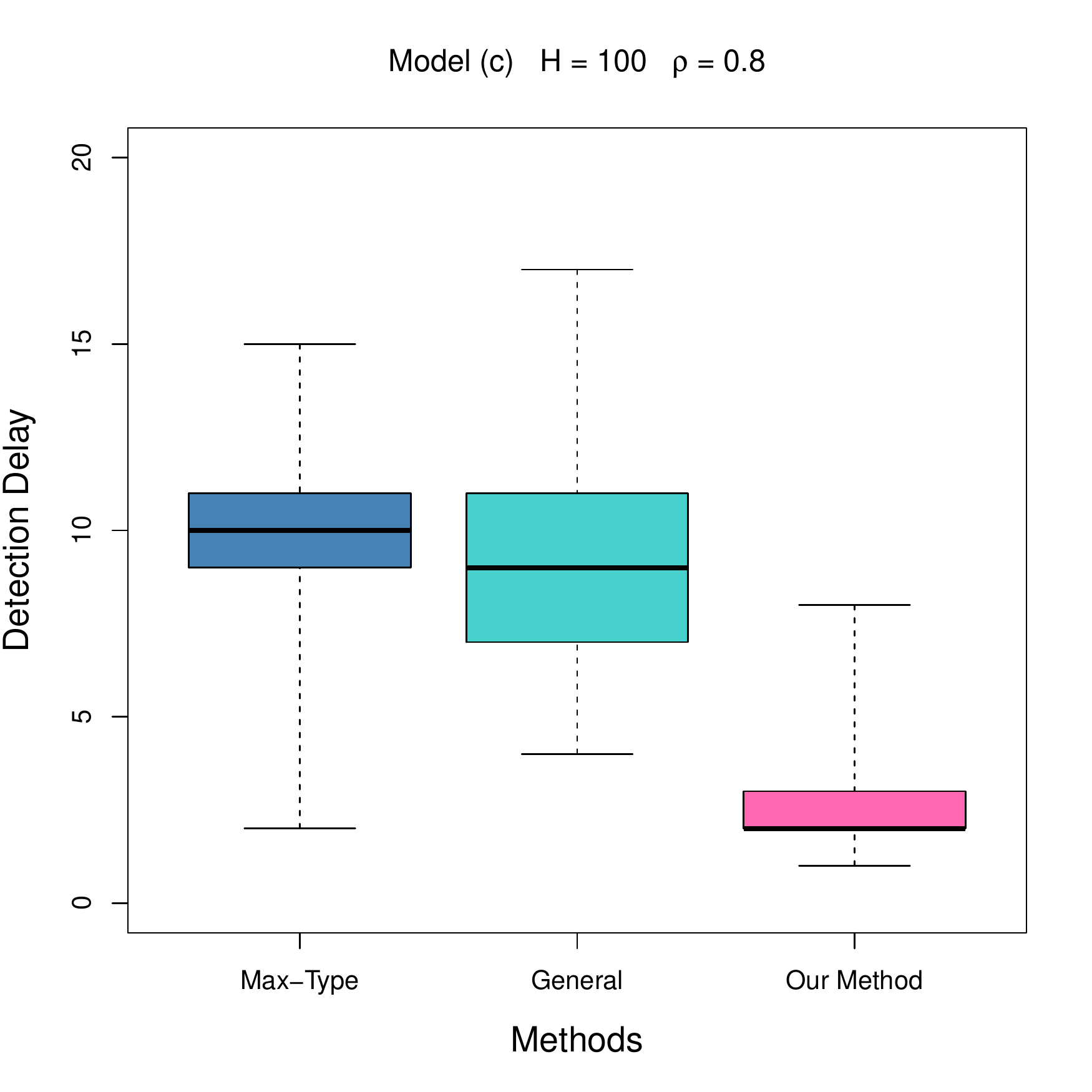}
\includegraphics[width=0.45\textwidth,height=0.45\textwidth]{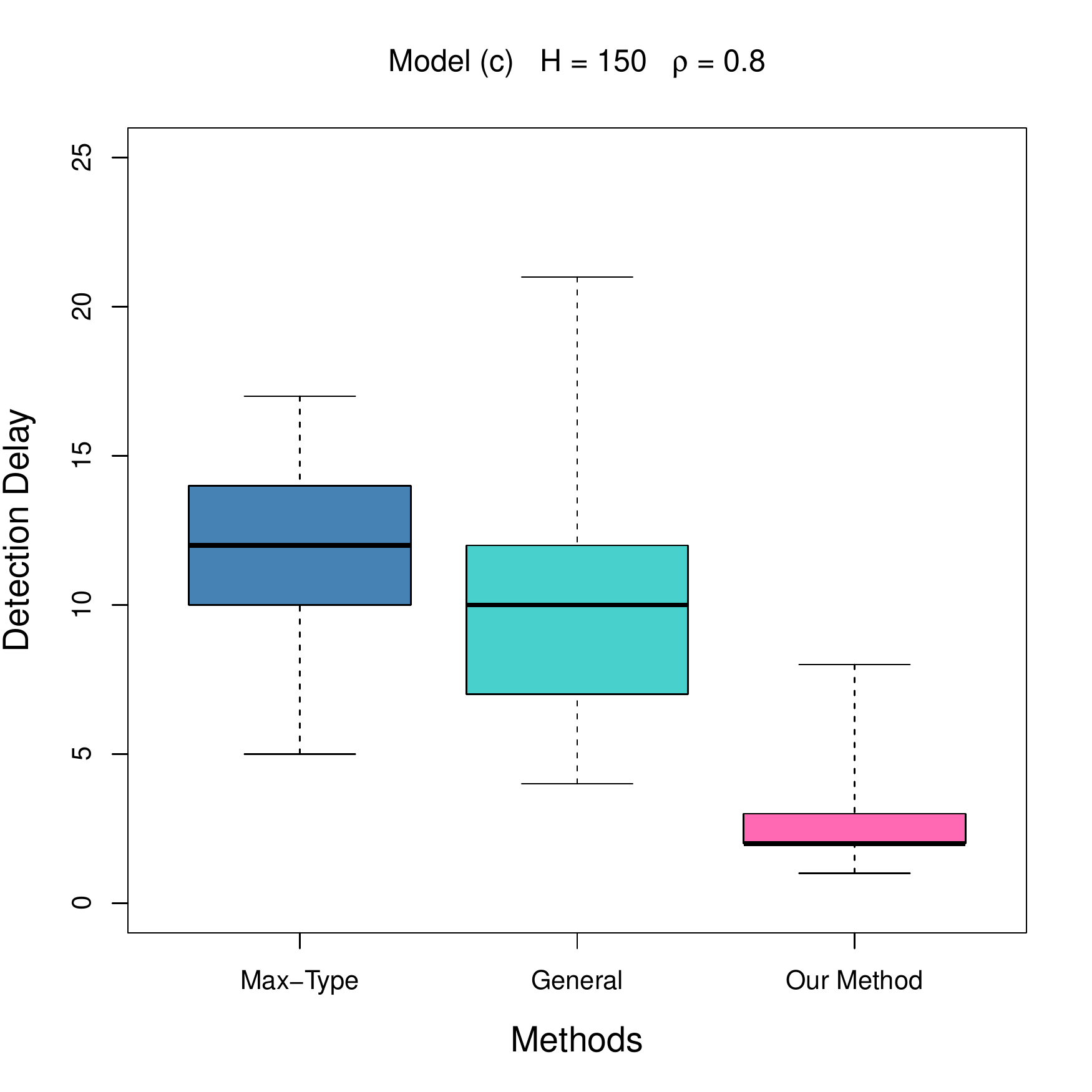}
\caption{Boxplots of EDDs for ``Max-type" and ``General" stopping rules in Chen (2019) and Chu and Chen (2018), and the proposed stopping rule. The results are based on $1000$ simulations under model (c). }
\label{comp3}
\end{center}
\end{figure}

We also compare the proposed stopping rule with some other stopping rules in the literature. Based on different edge-count statistics, Chen (2019) and Chu and Chen (2018) propose a series of stopping rules, among which the ones based on the generalized edge-count statistic and  based on the max-type edge-count statistic are more effective in detecting changes. The generalized and max-type stopping rules are based on a non-parameter framework that utilize nearest neighbor information among observations. The implementation of these two stopping rules are available in the R package {\it gStream}.  Similar to the authors, we choose a relatively larger nearest neighbors $5$ to gain more information. The ARL is specified at $5000$. Since they assume the observations are temporally independent, we consider $M=0$. Other setups are specified in the beginning of this section. Note that the stopping rules in Chen (2019) and Chu and Chen (2018) are proposed to detect the change point in distribution. When the change in distribution is indeed caused by the covariance structure, Figures \ref{comp1}-{\ref{comp3}} show that the proposed stopping rule performs better with much smaller EDDs than the two competitors.  %Moreover, the performance of the proposed stopping is more stable as its detection delays are more concentrated near the median.    

\subsection{Accuracy of the data-driven procedure for $M$}

\begin{figure}[t!]
\begin{center}
\includegraphics[width=0.4\textwidth,height=0.4\textwidth]{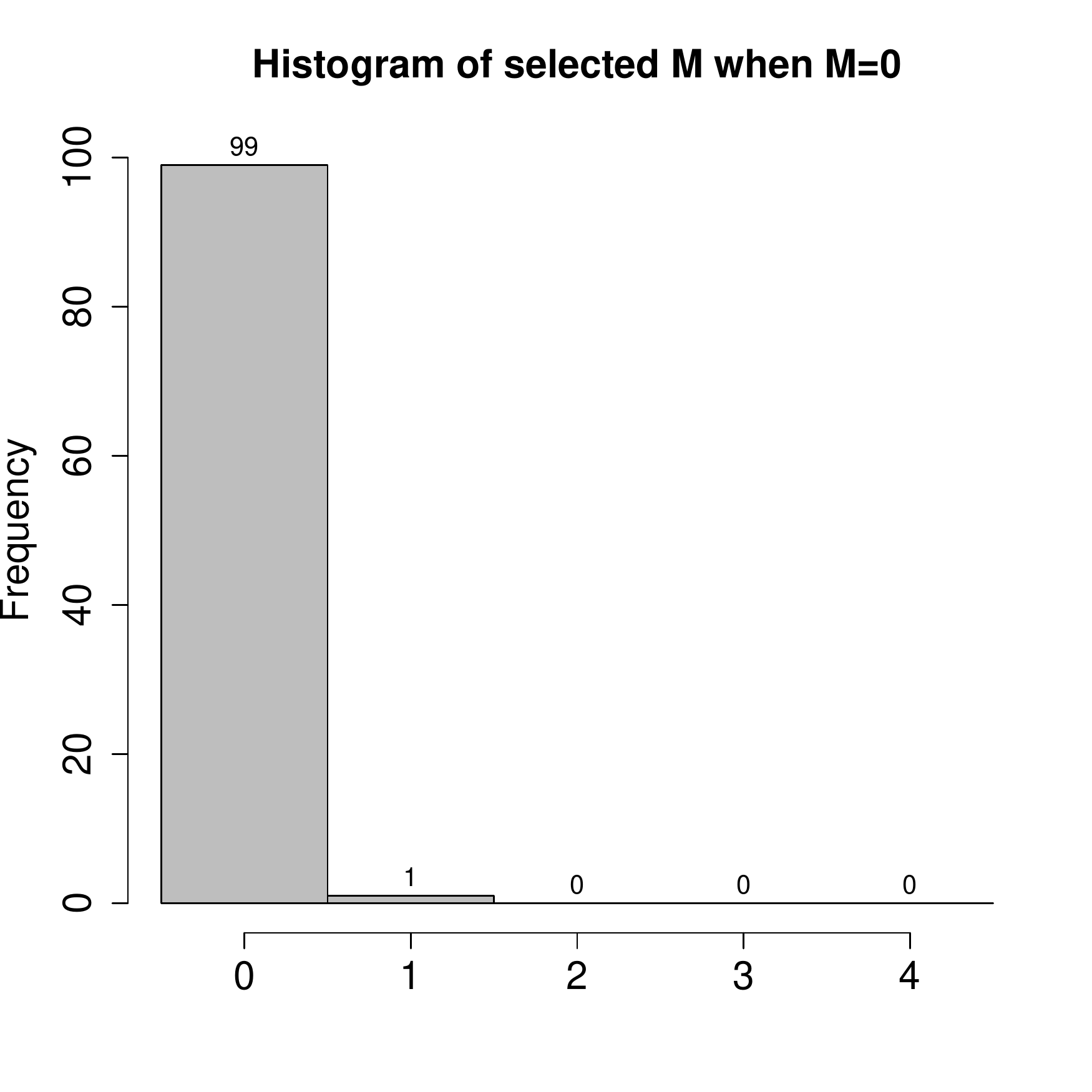}
\includegraphics[width=0.4\textwidth,height=0.4\textwidth]{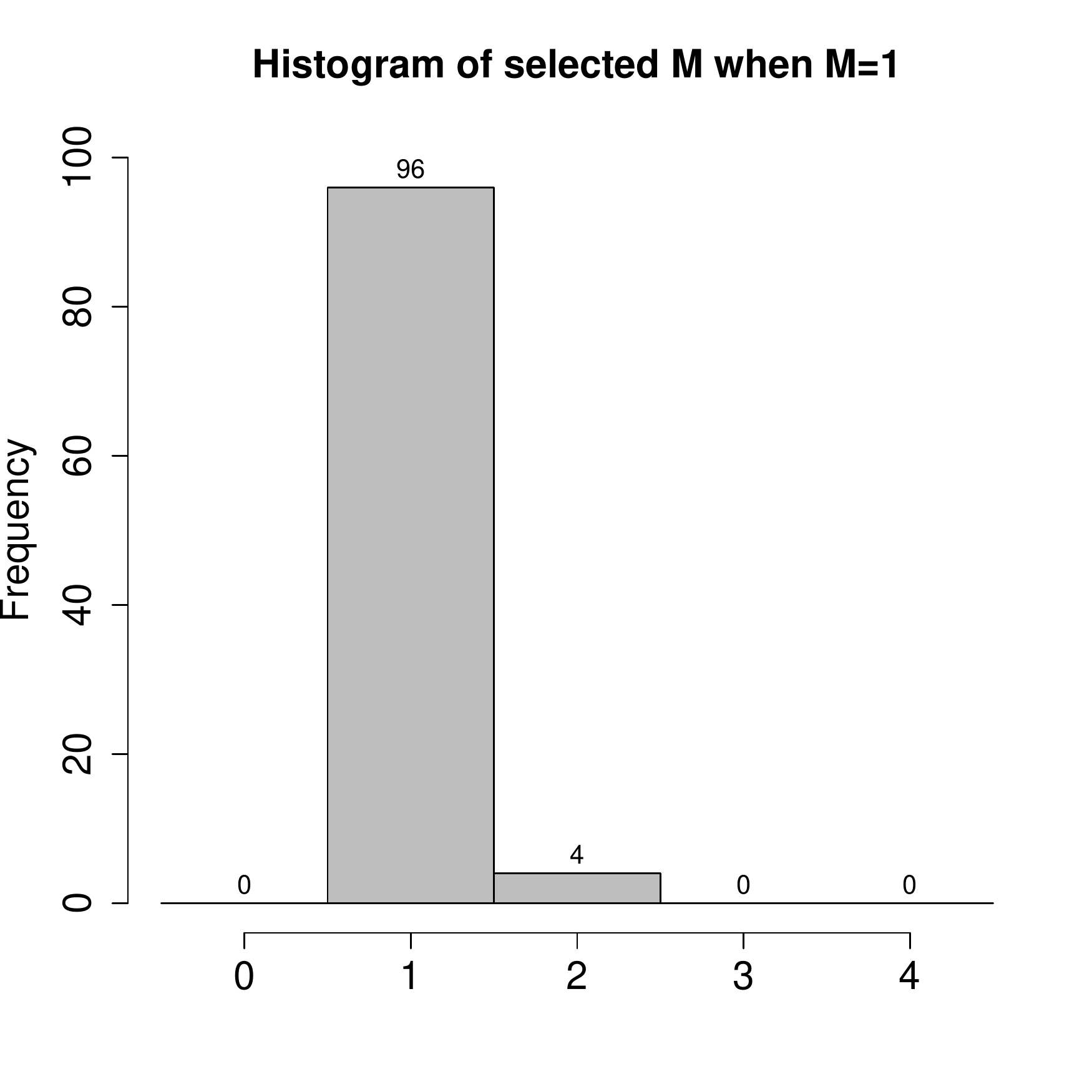}
\caption{Histograms of selected $M$ by the proposed data-driven procedure when the actual $M=0$ and $1$. The results are based on $100$ simulations.  }
\label{fig1}
\end{center}
\end{figure}

In the last part of simulation studies, we examine the data-driven procedure proposed in Section 3.4 for estimating $M$. For each simulation, a training sample of $200$ observations is generated from (\ref{data_g}) with $p=1000$. Figure \ref{fig1} illustrates the histograms of selected $M$ based on $100$ simulations when the actual $M=0$ and $1$. With $99$ and $96$ successes respectively,  the proposed data-driven procedure demonstrates its satisfactory performance for estimating the $M$.

\section{Case study}

Resting-state fMRI  is a method to explore brain's internal dynamic networks. We apply the proposed method to a resting-state fMRI  dataset obtained from the 2017 Human Connectome Project (HCP) data release. The data consist of 300 independent component analysis (ICA) component nodes ($p=300$) repeatedly observed over 1200 time points, collected for each of 1003 subjects. The publicly accessible dataset together with details about data acquisition and preprocessing procedures can be found in HCP website (http://www.humanconnectome.org).  

We detect the change in a real-time manner, in the sense that we pretend the observations in the dataset continually arrive in time. At each time, we determine whether the process should be terminated through the proposed stopping rule. Note that the proposed stopping is designed only for detecting the covariance change. When a detection process involves a change in the mean, it cannot be detected by the proposed stopping rule.
Despite such a limitation, we still apply the stopping rule to the dataset for the covariance change as the main interest of using the resting-state fMRI is to study the dynamic nature of brain connectivity (Cribben et al., 2013; Jeong et al., 2016). %we apply the proposed stopping rule to detect the covariance change by assuming a constant mean, even though the assumption may not be realistic.    

%If the termination occurs, it should be due to the covariance change rather than the mean change, as the proposed stopping rule is designed for the former by assuming stationarity of the latter. %On the other hand, if termination does not occur, it could be either  there is no change, or the change in the covariance is weaker than the minimum the stopping is able to detect, or the change is caused by other parameters such as the mean. 
%Since the main interest of using the resting-state fMRI is to study the dynamic nature of brain connectivity, we still apply the proposed stopping rule to detect the covariance change even if the assumption of constant mean may not be realistic.         

\begin{figure}[t!]
\begin{center}
\includegraphics[width=0.75\textwidth,height=0.45\textwidth]{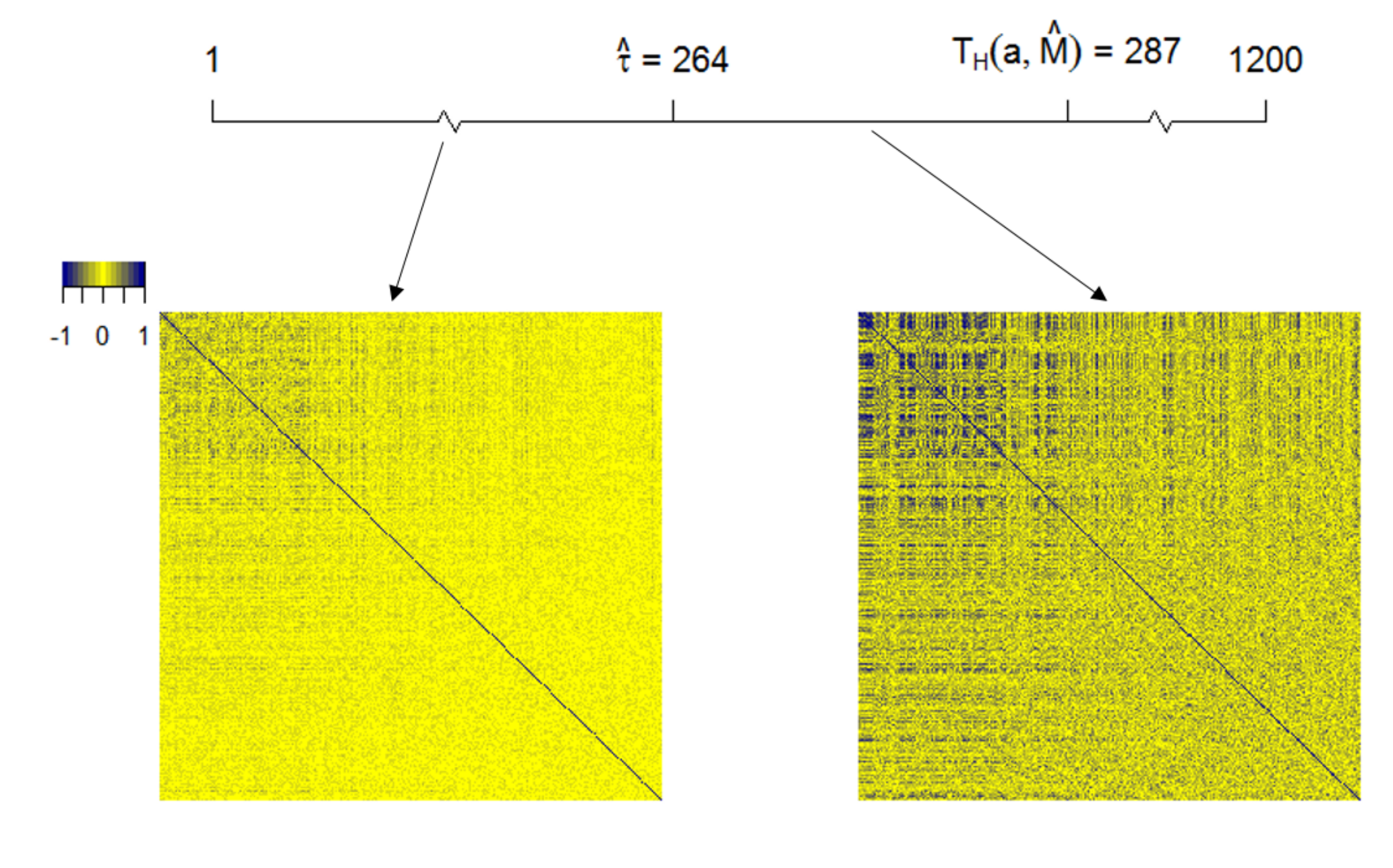}\\
\includegraphics[width=0.75\textwidth,height=0.45\textwidth]{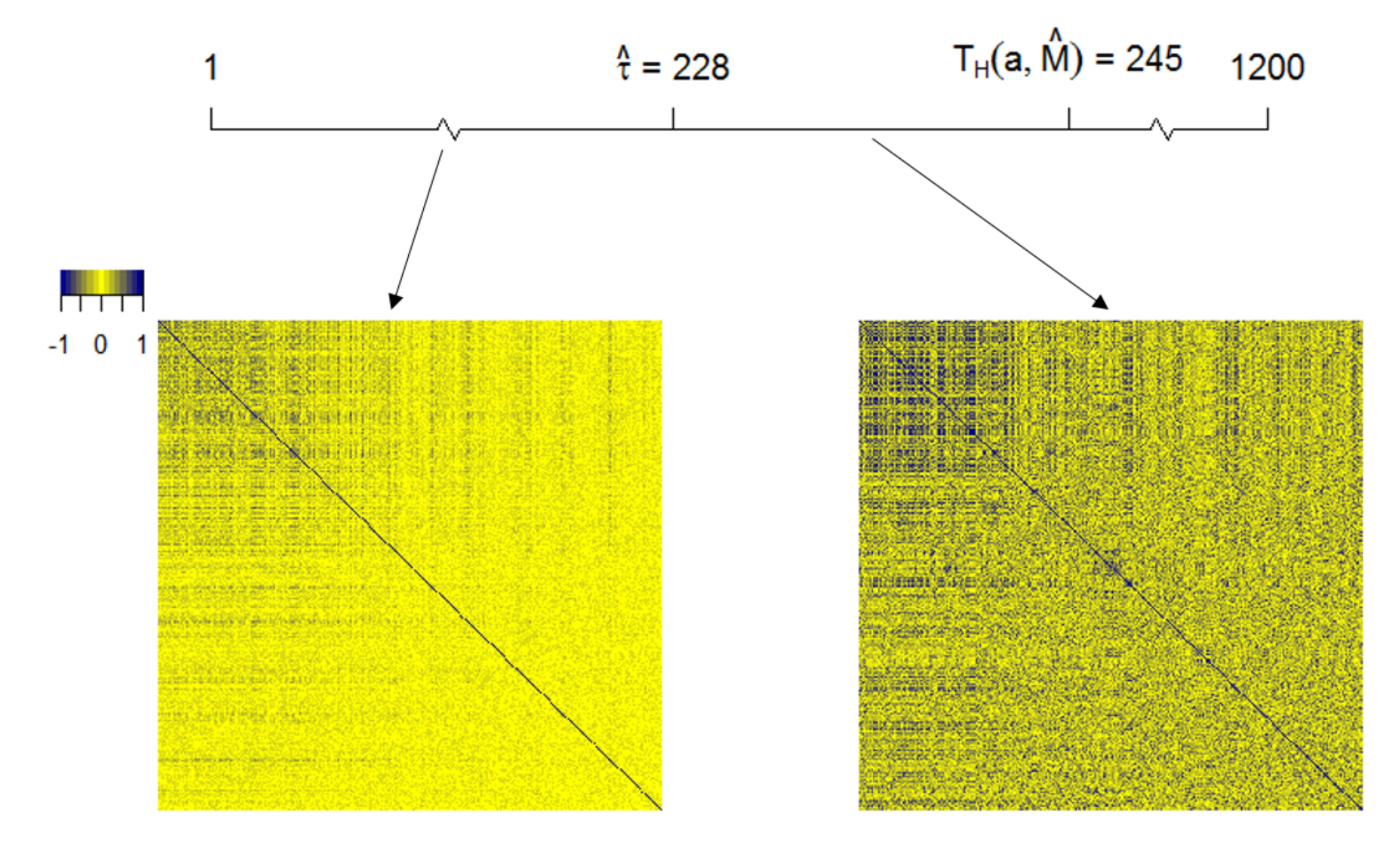}
\caption{Online change-point detection in the covariance structure of subject 103010 (upper panel) and subject 130417 (lower panel). Each panel illustrates the estimated correlation matrices before and after the estimated change point.}
\label{case-study}
\end{center}
\end{figure}

While there are 1003 subjects in the dataset, we randomly choose two subjects 103010 and 130417 to demonstrate the practical usefulness of the proposed method. The proposed stopping rule needs a training sample. We pretend that the first 200 observations of each time series are historical, and further justify their stationarity in the covariance structure through the testing procedure in Section 3.3. Here we use relatively large training sample size $200$ to attain precise estimation of nuisance parameters.   
%We further test the stationarity of the covariance structure in each of the two historical datasets is then tested by the procedure in Section 3.3, which leads to the conclusion that we fail to reject the null hypothesis. 
%we obtain the subject 103010 and subject 130417 with no change in the covariance of first 200 observations. We therefore focus on these two subjects with the training sample size $n_0=200$. 
Based on the training sample, we estimate $M$ by $5$ for the subject 103010 and $6$ for the subject 130417 using the method in Section 3.4 and obtain the sample mean $\hat{\mu}$ and the sample standard deviation of the test statistic using (\ref{var.est2}) in Section 2.3. Choosing the threshold $a=3.58$ so that the ARL is controlled around $5000$, we apply the proposed stopping rule with the window size $H=100$ and terminate the process at the time $287$ for the subject 103010 and the time $245$ for the subject 130417.

With each of the stopping times $287$ and $245$, we pull out the observations from time 1 to the stopping time and conduct some post analyses. 
The first analysis is change-point estimation.  Similar to Bai (2010), the change point is estimated by
$$\hat{\tau}=\arg\max_{1<t<T_H(a, \hat{M})}\hat{\mathcal{J}}_{t,\hat{M},H},$$
where $\hat{\mathcal{J}}_{t,\hat{M},H}$ is obtained by replacing $W_M(i,j)$ in $\hat{\mathcal{J}}_{n,M,H}$ with $A_{t,\hat{M}}(i,j)$ defined in (\ref{estimator}). The rationale of using the above estimator is that the expectation of $\hat{J}_{t,\hat{M},H}$ always attains its maximum at the true change point, as mentioned in Remark 2.2 of Section 2.2. %The difference between the stopped time point $t_s$ and the estimated change point $\hat{\tau}$, i.e. $t_s-\hat{\tau}$, gives a estimation of the detection delay.
The estimated change points are $264$ for the subject 103010 and $228$ for the subject 130417. With the two stopping times $287$ and $245$, the corresponding detection delays are $23$ for the subject 103010 and $17$ for the subject 130417, showing that the proposed stopping rule can quickly terminate the process when the brain's network change occurs.    

The second analysis is illustrating the actual change in the brain's network. 
For each subject, we estimate the correlation matrices before and after the estimated change point using the {\it glasso}. %To obtain the change in the two estimated correlation matrices, we take their difference and screen the noise through the similar procedure in Jeong et al. (2016).  
%The detection results for subject 103010 and subject 130417 are showed in Figure 1 and 2 respectively. The heatmaps in (a) of each figure present the estimated correlation matrices for each time partition determined by the estimated change point $\hat{\tau}$ and stopped time point $\tau_n$. To avoid the inference of data dependence in estimating the correlation matrix, the data within each partition were first divided in $k$ groups, and in each group, the indices of each data are at least $M$ apart. After estimating the correlation matrices in each group through graph lasso procedure, the final estimation of the correlation matrix, $\hat{R}_{g}$, was given by $\hat{R}_{g} = (1/k)\sum \hat{R}_{g,k}$. In (b) of each figure, the heatmap of the estimated changed connectivity are presented, the changed connectivity were estimated through the same procedure used in Jeong et al. (2016).
The obtained results for the two subjects are summarized in Figure \ref{case-study}, which clearly illustrates the brain's internal networks become stronger after the estimated change points. The results are consistent with recent studies that during the resting state, brain's networks activate when a subject focuses on internal tasks, and  exhibit dynamic changes within time scales of seconds to minutes (Allen et al. 2014; Calhoun et al. 2014; Chang and Glover 2010; Cribben et al. 2012; Handwerker et al. 2012; Hutchison et al. 2013b; Jeong et al. 2016; Monti et al. 2014). %that a distinct set of brain regions, known as default mode network, including Posterior Cingulate Cortex (PCC) and Medial Prefrontal Cortex (MPFC), shows significant functional connectivity during the resting state. Hereby, the empirical performance of the proposed method was demonstrated through case studies.

\section{Conclusion}

We propose a new procedure to detect the anomaly in the covariance structure of high-dimensional online data. The procedure is implementable when data are non-Gaussian, and involve both spatial and temporal dependence. We investigate its theoretical properties by deriving an explicit expression for the average run length (ARL) and an upper bound for the expected detection delay (EDD). The established ARL can be employed to obtain the level of the threshold in the stopping rule without running time-consuming Monte Carlo simulations. The derived upper bound demonstrates the impact of data dependence and magnitude of change in the covariance structure on the EDD. The theoretical properties are examined and justified by the empirical studies through both simulation and a real application. 

%The proposed stopping rule (\ref{stopping-rule}) is based on the sum-of-squares norm statistic. It is also interesting to consider a maximum norm based stopping rule
%\begin{eqnarray}
%T^*_H(b, M)= \mbox{inf} \biggl\{n-n_0:  \max_{n-M-H-2\le t \le n-M-2}\frac{\hat{\mathcal{J}}_{t, M, H}}{\hat{\sigma}_{\hat{\mathcal{J}}_{t, M, H}, 0}} > b, \,\, n > n_0 \biggr\}, \label{stopping-max}
%\end{eqnarray}
%where the statistic $\hat{\mathcal{J}}_{t, M, H}$ is obtained by replacing $W_M(i, j)$ in $\hat{\mathcal{J}}_{n, M, H}$ with $A_{t, M}(i, j)$ defined in (\ref{estimator}). In univariate settings, the sum-of-squares norm and the maximum num are employed respectively in two classical stopping rules:  the Shiryayev-Roberts procedure based on the sum of likelihood ratio statistics (Shiryayev 1963; Roberts 1966),  and the CUSUM procedure based on the maximum of likelihood ratio statistics (Page 1954; Lorden 1971). The two procedures and their properties have been thoroughly studied in the literature (Moustakides 1986; Pollak and Tartakovsky 2009). In high-dimensional settings, it will be interesting to study the impact of the two norms on the online change-point detection. It is thus worth deriving the ARL and EDD of the maximum norm based stopping rule (\ref{stopping-max}) and comparing them with those of the sum-of-squares norm based stopping rule (\ref{stopping-rule}). We leave this for future study.

\section{Appendix: Technical Details}

\subsection{Proof of Proposition 1}

From (\ref{estimator}), $X_i$ and $X_j$ in $\hat{\mathcal{J}}_{n, M}$ are $M$ apart because of the indicator function in $W_M(i, j)$. Using (C1), we see that $X_i$ and $X_j$ are independent. As a result, $\mbox{E}(\hat{\mathcal{J}}_{n, M})=n^{-2} \sum_{i,j} W_M(i, j)\mbox{tr}(\Sigma_i \Sigma_j)$ by the model (\ref{model}), which gives the expectation under the alternative hypothesis. Specially, under the null hypothesis,
\[
\mbox{E}(\hat{\mathcal{J}}_{n, M})=\frac{\mbox{tr}(\Sigma^2)}{n^2}\sum_{i,j} W_M(i, j)=0,
\]
because $\sum_{i,j} W_M(i, j)=0$. This completes the proof of Proposition 1.

\subsection{Proof of Proposition 2}

Note that $\V(\hat{\mathcal{J}}_{n, M})=\E(\hat{\mathcal{J}}_{n, M}^2)-\E^2(\hat{\mathcal{J}}_{n, M})$, where $\mbox{E}^2(\hat{\mathcal{J}}_{n, M})=n^{-4} \sum_{i,j} \sum_{k,l} W_M(i, j) W_M(k, l) \mbox{tr}(\Sigma_i \Sigma_j)\mbox{tr}(\Sigma_k \Sigma_l)$ from Proposition 1.
We thus only need to derive $\mbox{E}(\hat{\mathcal{J}}_{n, M}^2)$, which, from (\ref{model}) and (C1), is
\begin{eqnarray}
\mbox{E}(\hat{\mathcal{J}}^2_{n, M})&=&\frac{1}{n^4} \sum_{i,j}\sum_{k,l} W_M(i, j) W_M(k, l) \mbox{E}(X_i^T X_j X_j^T X_i X_k^T X_l X_l^T X_k)\nonumber\\
&=&\frac{1}{n^4} \sum_{i,j}\sum_{k,l} W_M(i, j) W_M(k, l) \mbox{tr}(\Sigma_i \Sigma_j) \mbox{tr}(\Sigma_k \Sigma_l)\nonumber\\
&+&\frac{4}{n^4} \sum_{i,j}\sum_{k,l} W_M(i, j) W_M(k, l)\mbox{tr}^2\{C(i-k)C(l-j)\}\nonumber\\
&+&\frac{1}{n^4} \sum_{i,j}\sum_{k,l} W_M(i, j) W_M(k, l)
\bigg[16\mbox{tr}\{\Sigma_lC(i-k)\Sigma_jC(k-i)\}\nonumber\\
&+&4\mbox{tr}\{C(k-i)C(j-l)C(k-i)C(j-l)\} \nonumber\\
&+&8\beta \mbox{tr}(\Gamma_i^T\Gamma_j \Gamma_j^T \Gamma_i \circ \Gamma_k^T\Gamma_l \Gamma_l^T \Gamma_k)+8\beta \mbox{tr}(\Gamma_i^T\Gamma_j \Gamma_k^T \Gamma_l \circ \Gamma_i^T\Gamma_j \Gamma_k^T \Gamma_l)\nonumber\\
&+&\sum_{m}2\beta^2 \mbox{tr}(\Gamma_j^T\Gamma_i e_m e_m^T \Gamma_k^T \Gamma_l \circ \Gamma_j^T\Gamma_i e_m e_m^T \Gamma_k^T \Gamma_l)\biggr], \nonumber
\end{eqnarray}
where for any square matrices $A$ and $B$, the symbol $A \circ B=(a_{ij} b_{ij})$, and $e_m$ is the unit vector with the only non-zero element at the $m$th component. By applying (C2) and subtracting $\mbox{E}^2(\hat{\mathcal{J}}_{n, M})$ in Proposition 1, we have
\[
\mbox{E}(\hat{\mathcal{J}}^2_{n, M})=\frac{4}{n^4} \sum_{i,j}\sum_{k,l} W_M(i, j) W_M(k, l)\mbox{tr}^2\{C(i-k)C(l-j)\}\{1+o(1)\}.
\]
This completes the proof of Proposition 2.

\subsection{Proof of Theorem 1}

We need to derive the cumulative distribution function of $T_H(a, M)$. From (\ref{stopping-rule}),
\[
\mbox{P}_{\infty}\{T_H(a, M)\le t \}=\mbox{P}_{\infty} \biggl(\max_{0 \le i \le t} \left | \frac{\hat{\mathcal{J}}_{n_0+i, M, H}}{\hat{\sigma}_{{n_0, M, H}}} \right | > a \biggr).
\]
The cumulative distribution function of $T_H(a, M)$ thus depends on the distribution of
${\hat{\mathcal{J}}_{n_0+i, M, H}}/{\hat{\sigma}_{{n_0, M, H}}}$, which will be shown to converge to a stationary Gaussian process.

%Letting $q=i/t$, we want to show that ${\hat{\mathcal{J}}_{n_0+i, M, H}}/{\hat{\sigma}_{\hat{\mathcal{J}}_{n_0+i, M, H}, 0}}$ converges to a stationary Gaussian process $Z(q)$ as $t \to \infty$. As a result,
%\[
%\mbox{P}_{\infty}\{T_H(a)\le t \}=1-\mbox{P}_{\infty} \biggl\{\max_{0 \le q \le 1} \left | Z(q) \right | \le a \biggr\}.
%\]
%where $Q=t/H$.

To simplify notation, we let $\hat{\mathcal{J}}_{n_0+i, M, H} \equiv \hat{\mathcal{J}}_{i,M}$, and $\hat{\sigma}_{{n_0, M, H}} \equiv \hat{\sigma}_0$. The Gaussian process can be established by showing (i) the joint asymptotic normality of $(\hat{\sigma}_0^{-1}\hat{\mathcal{J}}_{i_1,M},\dots,\hat{\sigma}_0^{-1}\hat{\mathcal{J}}_{i_d,M})'$ for any $i_1 < i_2< \cdots < i_d$. (ii) the tightness of $\hat{\sigma}_0^{-1}\hat{\mathcal{J}}_{i,M}$. To prove (i), we apply the Cram${\rm\acute{e}}$r-Wold device to show that for any non-zero $a_1, \cdots, a_d$, $\sum_{l=1}^d \hat{\sigma}_0^{-1} a_l\hat{\mathcal{J}}_{i_l,M}$ is asymptotic normal. Since the proof is similar to that of Theorem 3, we omit it. We thus only need to prove (ii).

Toward this end, we first obtain the leading order of $\mbox{Var}(\hat{\mathcal{J}}_{i,M})$, which is
\begin{eqnarray*}
\mbox{Var}(\hat{\mathcal{J}}_{i,M})
&=&\frac{4}{H^4}\sum_{i, j=1}^H \sum_{h_1,h_2}W_M(i, j) W_M(i-h_1, j+h_2)\mbox{tr}^2\{C(h_1)C(h_2)\}\\
&=&\frac{4}{H^4}\sum_{h_1,h_2}\mbox{tr}^2\{C(h_1)C(h_2)\}\bigg(\frac{6\pi^2-51}{18}H^4\bigg).
%&=&\frac{4}{H^4}\sum_{h_1,h_2}\mbox{tr}^2\{C(h_1)C(h_2)\}(C_{\pi}H^4)\{1+o(1)\}.
\end{eqnarray*}

Let $i_1,i_2\in\{1,\dots,t\}$. Next, we derive the leading order of $\mbox{Cov}(\hat{\mathcal{J}}_{i_1,M}, \hat{\mathcal{J}}_{i_2,M})$, which equals $\mbox{E}(\hat{\mathcal{J}}_{i_1,M} \hat{\mathcal{J}}_{i_2,M})$ when there is no any change point.
Let $i_d\equiv i_2-i_1$.

For $i_d\in\{1,\dots,H-1\}$ and $H-i_d=O(H)$, under (C1), the leading order of $\mbox{Cov}(\hat{\mathcal{J}}_{i_1,M}, \hat{\mathcal{J}}_{i_2,M})$ depends on $i_d$ can be derived to be
\begin{eqnarray*}
\mbox{Cov}(\hat{\mathcal{J}}_{i_1,M},\hat{\mathcal{J}}_{i_2,M})
=\frac{4}{H^4}\sum_{h_1,h_2}\mbox{tr}^2\{C(h_1)C(h_2)\}
\bigg\{\frac{6\pi^2-51}{18}(H-i_d)^4\bigg\}.
\end{eqnarray*}

For $i_d\in\{1,\dots,H-1\}$ and $H-i_d=o(H)$, or for $i_d\geq H$, ${\rm Cov}(\hat{\mathcal{J}}_{i_1,M},\hat{\mathcal{J}}_{i_2,M})$ can be shown is the smaller order of ${\rm Var}(\hat{\mathcal{J}}_{i,M})$, i.e.
$$
\mbox{Cov}(\hat{\mathcal{J}}_{i_1,M},\hat{\mathcal{J}}_{i_2,M})
=o\bigg[\frac{4}{H^4}\sum_{h_1,h_2}\mbox{tr}^2\{C(h_1)C(h_2)\}\bigg(\frac{6\pi^2-51}{18}H^4\bigg)\bigg].
$$

We want to show the tightness of ${\sigma}_{0}^{-1}\hat{\mathcal{J}}_{i,M}$. Then the tightness of $\hat{\sigma}_{0}^{-1}\hat{\mathcal{J}}_{i,M}$ can be established by the Slutsky theorem because $\hat{\sigma}_{0}$ is ratio-consistent to ${\sigma}_{0}$ according to Theorem 3. Consider $i=q^*\cdot t$, for $q^*=i/t\in(0,1)$, with $i=1,\dots,t$. It is equivalent to show the tightness of $G(i/t)$, where $G(i/t)=G(q^*)=\sigma_{0}^{-1}\hat{\mathcal{J}}_{i,M}$. For $0<q^*<r^*<1$,
\begin{eqnarray*}
\mbox{E}|G(r^*)-G(q^*)|^2&=&\sigma_{0}^{-1}\mbox{E}|\hat{\mathcal{J}}_{i_1,M}-\hat{\mathcal{J}}_{i_1,M}|^2\\
&=&\sigma_{0}^{-1}\{\mbox{E}(\hat{\mathcal{J}}_{i_1,M}^2)+\mbox{E}(\hat{\mathcal{J}}_{i_2,M}^2)
-2\mbox{E}(\hat{\mathcal{J}}_{i_1,M}\hat{\mathcal{J}}_{i_2,M})\}.
\end{eqnarray*}
When there is no any change point,
\begin{eqnarray*}
&&\mbox{E}(\hat{\mathcal{J}}_{i_1,M}^2)=\mbox{E}(\hat{\mathcal{J}}_{i_2,M}^2)=\mbox{Var}(\hat{\mathcal{J}}_{i,M})\\
&=&\frac{4}{H^4}\sum_{h_1,h_2}\mbox{tr}^2\{C(h_1)C(h_2)\}\bigg(\frac{6\pi^2-51}{18}H^4\bigg)\{1+o(1)\}.
\end{eqnarray*}
For any $i_1,i_2\in\{1,\dots,t\}$, and $i_2-i_1=i_d\in\{1,\dots,H-1\}$, as $H \to \infty$,
\begin{eqnarray*}
\mbox{E}|G(r^*)-G(q^*)|^2
&\leq& C\frac{(4/H^4)\sum_{h_1,h_2}\mbox{tr}^2\{C(h_1)C(h_2)\}2\{H^4-(H-i_d)^4\}}{(4/H^4)\sum_{h_1,h_2}\mbox{tr}^2\{C(h_1)C(h_2)\}H^4}\\
&\leq&C\bigg(\frac{i_d}{H}\bigg).
\end{eqnarray*}
For $i_d\geq H$,
\begin{eqnarray*}
\mbox{E}|G(r^*)-G(q^*)|^2
\leq\frac{C(4/H^4)\sum_{h_1,h_2}\mbox{tr}^2\{C(h_1)C(h_2)\}2\{H^4+o(H^4)\}}{(4/H^4)\sum_{h_1,h_2}\mbox{tr}^2\{C(h_1)C(h_2)\}H^4}
\leq C.
\end{eqnarray*}
Therefore, by Chebyshev's inequality, if $1\leq i_d\leq H-1$,
\begin{eqnarray*}
\mbox{P}(|G(r^*)-G(q^*)|\geq \lambda)\leq \frac{\mbox{E}|G(r^*)-G(q^*)|^2}{\lambda^2}
\leq(C/\lambda^2)(i_d/H).
\end{eqnarray*}
Let $H/t=d$, then
$$i_d/H=(i_2-i_1)/H=(r^*-q^*)t/H=(r^*-q^*)/d,$$
and $\{(r^*-q^*)/d\}\in(0,1)$.
If $i_d\geq H$, or equivalently $\{(r^*-q^*)/d\}\geq1$,
\begin{eqnarray*}
\mbox{P}(|G(r^*)-G(q^*)|\geq \lambda)\leq \frac{\mbox{E}|G(r^*)-G(q^*)|^2}{\lambda^2}
\leq C/\lambda^2.
\end{eqnarray*}

Let
\[ f_d\{(q^*,r^*]\}=\begin{cases}
(r^*-q^*)/d,\quad if\quad r^*-q^*< d\\
1,\quad if\quad r^*-q^*\geq d,
\end{cases}\]
then
$$\mbox{P}(|G(r^*)-G(q^*)|\geq \lambda)\leq (C/\lambda^2)f_d\{(q^*,r^*]\}.$$
Let $\xi_{i}=G(i/t)-G\{(i-1)/m\}$, for $i=1,\dots,t.$
Then $S_{i}=\xi_1+\cdots+\xi_{i}=G(i/t)$ with $S_0=0$.
Therefore,
$$\mbox{P}(|S_{i_2}-S_{i_1}|\geq \lambda)\leq(C/\lambda^2)f_d\{(q^*,r^*]\}.$$
For any $0<p^*<q^*<r^*<1$, $G(p^*)=S_{i_0}$, $G(q^*)=S_{i_1}$ and $G(r^*)=S_{i_2}$, respectively.
Let $m^*=|G(q^*)-G(p^*)|\wedge|G(r^*)-G(q^*)|$. Then
\begin{eqnarray*}
\mbox{P}(m^*\geq \lambda)
&=&\mbox{P}\big[\{|G(q^*)-G(p^*)|\geq \lambda\}\cap\{|G(r^*)-G(q^*)|\geq \lambda\}\big]\\
&\leq&\mbox{P}^{1/2}(|S_{i_1}-S_{i_0}|\geq \lambda)\cdot \mbox{P}^{1/2}(|S_{i_2}-S_{i_1}|\geq \lambda)\\
&\leq&(C/\lambda)f_d^{1/2}\{(p^*,q^*]\}(C/\lambda)f_d^{1/2}\{(q^*,r^*]\}\\
&\leq&(C/\lambda^2)[f_d\{(p^*,q^*]\}+f_d\{(q^*,r^*]\}].
\end{eqnarray*}
If $q^*-p^*<d$ and $r^*-q^*<d$, or equivalently $r^*-p^*<2d$,
\begin{eqnarray*}
\mbox{P}(m^*\geq\lambda)
\leq(C/\lambda^2)\bigg\{\frac{q^*-p^*}{d}+\frac{r^*-q^*}{d}\bigg\}
\leq(C/\lambda^2)\bigg(\frac{r^*-p^*}{d}\bigg).
\end{eqnarray*}
If $q^*-p^*<d$ and $r^*-q^*\geq d$, or $q^*-p^*\geq d$ and $r^*-q^*< d$, but $r^*-p^*<2d$,
\begin{eqnarray*}
\mbox{P}(m^*\geq \lambda)&\leq&(C/\lambda^2)\bigg\{\frac{q^*-p^*}{d}+1\bigg\}\\
&\leq&(C/\lambda^2)\bigg(\frac{q^*-p^*}{d}+\frac{r^*-q^*}{d}\bigg)
\leq (C/\lambda^2)\bigg(\frac{r^*-p^*}{d}\bigg).
\end{eqnarray*}
If $q^*-p^*<d$ and $r^*-q^*\geq d$, or $q^*-p^*\geq d$ and $r^*-q^*< d$, but $r^*-p^*\geq2d$,
\begin{eqnarray*}
\mbox{P}(m^*\geq \lambda)&\leq&(C/\lambda^2)\bigg\{\frac{q^*-p^*}{d}+1\bigg\}\leq 2C/\lambda^2.
\end{eqnarray*}
If $q^*-p^*\geq d$ and $r^*-q^*\geq d$, and $r^*-p^*\geq2d$,
\begin{eqnarray*}
\mbox{P}(m^*\geq \lambda)
\leq 2C/\lambda^2.
\end{eqnarray*}
Let
\[ \mu_{\alpha,d}\{(p^*,r^*]\}=\begin{cases}
(\frac{r^*-p^*}{d})^{\frac{1}{2\alpha}},\quad if\quad r^*-p^*< 2d\\
2^{\frac{1}{2\alpha}},\quad if\quad r^*-p^*\geq 2d,
\end{cases}\]
where $\alpha>\frac{1}{2}$. Then $\mu_{\alpha,d}\{(p^*,r^*]\}$ is a finite measure on $T=(0,1]$. For any $\epsilon>0$ and $p^*,q^*,r^*\in T=(0,1]$,
$$\mbox{P}(m^*\geq \lambda)\leq (C/\lambda^2)\mu_{\alpha,d}^{2\alpha}\{(p^*,r^*]\}.$$
Let
\begin{eqnarray*}
L(G)=\sup_{p^*\leq q^*\leq r^*}m^*=\max_{i_0\leq i_1\leq i_2}|S_{i_1}-S_{i_0}|\wedge|S_{i_2}-S_{i_1}|.
\end{eqnarray*}
Using Theorem 10.3 in Billingsley (1999), we conclude
$$P\{L(G)\geq\lambda\}\leq \frac{KC}{\lambda^2}\mu_{\alpha,d}^{2\alpha}\{(0,1]\},$$
where $K$ is a constant. As $t\gg H$, $d=H/t$ is close to zero, and $2d<(1-0)$. Hence, $\mu_{\alpha,d}^{2\alpha}\{(0,1]\}=2$, and
$$P\{L(G)\geq\lambda\}\leq \frac{2KC}{\lambda^2}.$$
From (10.4) in Billingsley (1999), we obtain
$$\max_{1\leq i\leq t}|S_{i}|\leq L(G)+|S_t|.$$
Since
$\mbox{E}|S_t|^2=\sigma_{0}^{-2}\mbox{E}(\hat{\mathcal{J}}_{t,M}^2)=1$,
we have
\begin{eqnarray*}
\mbox{P}(\max_{1\leq i\leq t}|S_{i}|\geq\lambda)
&\leq& \mbox{P}\bigg\{L(G)\geq\frac{1}{2}\lambda\bigg\}+P\bigg(|S_t|\geq\frac{1}{2}\lambda\bigg)\\
&\leq&\frac{2KC}{(\frac{1}{2}\lambda)^2}+\frac{\mbox{E}|S_t|^2}{(\frac{1}{2}\lambda)^2}\leq\frac{KC}{\lambda^2}.
\end{eqnarray*}
If $\lambda$ goes to infinity, the above probability converges to zero. Therefore, $S_{i}$ is tight or equivalently $\hat{\mathcal{J}}_{i,M}/\sigma_{0}$ is tight.

Let $q=i/H$ and let $Y(q)=Y(i/H)\equiv\hat{\mathcal{J}}_{i,M}/\sigma_{0}$. For $0\leq p\leq q$, consider $|p-q|\to 0$, then we have, as $H\to\infty$,
\begin{eqnarray*}
|p-q|\to 0\Rightarrow |i_1-i_2|/H\to 0\Rightarrow i_d/H\to 0\Rightarrow i_d=o(H).
\end{eqnarray*}
If $i_d=o(H)$,
\begin{eqnarray*}
\mbox{Cov}\{Y(p),Y(q)\}&=&\sigma_{0}^{-2}\mbox{E}(\hat{\mathcal{J}}_{i_1,M}\hat{J}_{i_2,M})\\
&=&\frac{(4/H^4)\sum_{h_1,h_2}\mbox{tr}^2\{C(h_1)C(h_2)\}\{(H-i_d)^4\}}{(4/H^4)\sum_{h_1,h_2}\mbox{tr}^2\{C(h_1)C(h_2)\}H^4}\{1+o(1)\}\\
&=&\{(H-i_d)^4/H^4\}\{1+o(1)\}=1-4(i_d/H)+o\{(i_d/H)\}\\
&=&1-4|p-q|+o\{|p-q|\}.
\end{eqnarray*}
On the other hand, if $|p-q|\to\infty$ or $i_d/H\to\infty$, $\mbox{Cov}\{Y(p),Y(q)\}=0$.

As a result, $\{Y(q),q\geq 0\}$ converges to $\{Z(q),q\geq 0\}$, which is a stationary Gaussian process with zero mean, unit variance and covariance function of the form
$$r(|p-q|)=\mbox{Cov}\{Z(p),Z(q)\}=1-4|p-q|+o(|p-q|),$$
as $|p-q|\to 0$. On the other hand, as $|p-q|\to\infty$, $r(|p-q|)\log(|p-q|)\to 0$.

%Moreover, we can show that the stationary Gaussian process $\{ Z(q): q \ge 0\}$ has zero mean and the variance-covariance function
%\[
%\mbox{Cov}\{Z(s), Z(t)\}=1- 4|s-t|+ o(|s-t|).
%\]
Let $Q=t/H$. From Finch (2003), as $Q \to \infty$, $\max_{0 \le q \le Q} \left | Z(q) \right |$ has the Gumbel distribution so that
\[
\mbox{P}_{\infty} \biggl\{\max_{0 \le q \le Q} \left | Z(q) \right | \le a \biggr\} = \mbox{exp}\biggl[-2 \mbox{exp}\biggl\{g(t/H, a)\biggr\}\biggr],
\]
where
\[
g(t/H, a)={2\log (t/H)}+1/2 \log \log (t/H)+\log (4/\sqrt{\pi})-a\sqrt{2\log (t/H)}.
\]
As a result, when $t > H$, 
\[
\mbox{P}_{\infty}\{T_H(a, M)\le t \}= 1-\mbox{exp}\biggl[-2 \mbox{exp}\biggl\{g(t/H, a)\biggr\}\biggr].
\]

When $t=H$ and as $H \to \infty$, 
\[
\mbox{P}_{\infty} \biggl\{\max_{0 \le q \le 1} \left | Z(q) \right | \le a \biggr\}= \mbox{exp}\biggl\{-(2\sqrt{\pi})^{-1} H \mbox{exp}(-a^2/2)\biggr\},
\]
which has the order of $1-1/(2\sqrt{\pi}) H \mbox{exp}(-a^2/2)$ because $H=o\{\mbox{exp}(a^2/2) \}$. As a result,
\[
\mbox{P}_{\infty}\{T_H(a, M)\le H \}= 1/(2\sqrt{\pi}) H \mbox{exp}(-a^2/2),
\]
which decays to zero as $H=o\{\mbox{exp}(a^2/2) \}$.

We next derive the expectation of $T_H(a, M)$. Since the support of $T_H(a, M)$ is non-negative, we have
\[
\mbox{E}_{\infty}\{T_H(a, M) \}= \int_0^{\infty} \{1- F_{T_H(a, M)} (t)\} dt,
\]
where $F_{T_H(a, M)} (t)$ is the cumulative distribution function of $T_H(a, M)$ evaluated at $t$. From the above results, we have
%Since we have already derived the cumulative distribution function when $t > H$, we then have
\begin{eqnarray}
\mbox{E}_{\infty}\{T_H(a, M) \}&=& \int_0^{H} \{1- F_{T_H(a, M)} (t)\} dt+\int_H^{\infty} \{1- F_{T_H(a, M)} (t)\} dt\nonumber\\
&=& \biggr(H + \int_H^{\infty}\mbox{exp}\biggl[-2 \mbox{exp}\biggl\{g(t/H, a)\biggr\}\biggr] dt\biggr)\{1+o(1)\}. \nonumber
\end{eqnarray}
This completes the proof of Theorem 1.

\subsection{Proof of Theorem 2}

We first prove that the supremum of the EDDs attains at the immediate change point $\tau=n_0$. Equivalently, we need to show that for any $\tau > n_0$,  
\[
\mbox{E}_{\tau}\{T_H(a, M)-(\tau-n_0)| T_H(a, M)>\tau-n_0\} \le \mbox{E}_0\{T_H(a, M)\}.
\]

To simplify notation, we let $\tau^*=\tau-n_0$ and $T^*=T_H(a, M)-\tau^*$. Then to show the above inequality, we only need to show that 
\[
\mbox{E}_{\tau}\{T^*| T^*>0\} \le \mbox{E}_0\{T^*\}.
\]
Since 
\[
\mbox{E}_{\tau}\{T^*| T^*>0\}=\int_0^{\infty} \{1-\mbox{P}_{\tau}(T^*<t| T^*>0)\} dt, \quad \mbox{and}
\]
\[
\mbox{E}_{0}\{T^*\}=\int_0^{\infty} \{1-\mbox{P}_{0}(T^*<t)\} dt, 
\]
we only need to show that
\begin{eqnarray}
\mbox{P}_{\tau}(T^*<t| T^*>0)\ge \mbox{P}_{0}(T^*<t), \label{ine} 
\end{eqnarray}

First, the probability on the left hand side of (\ref{ine}) is
\begin{eqnarray}
\mbox{P}_{\tau}(T^*<t| T^*>0)=\frac{\mbox{P}_{\tau}(T^*<t)-\mbox{P}_{\tau}(T^*<0) }{1-\mbox{P}_{\tau}(T^*<0)}, \label{eq2} 
\end{eqnarray}
where 
\[
\mbox{P}_{\tau}\{T^*< t \}=\mbox{P}_{\tau} \biggl(\max_{0 \le i \le t+\tau^*} \left | \frac{\hat{\mathcal{J}}_{n_0+i, M, H}}{\hat{\sigma}_{{n_0, M, H}}} \right | > a \biggr), \quad \mbox{and} 
\]
\[
\mbox{P}_{\tau}\{T^*< 0 \}=\mbox{P}_{\tau} \biggl(\max_{0 \le i \le \tau^*} \left | \frac{\hat{\mathcal{J}}_{n_0+i, M, H}}{\hat{\sigma}_{{n_0, M, H}}} \right | > a \biggr). 
\]
From the above two probabilities, we can define two events
\[
A=\{\max_{0 \le i \le t+\tau^*} \left | \frac{\hat{\mathcal{J}}_{n_0+i, M, H}}{\hat{\sigma}_{{n_0, M, H}}} \right | > a \}, \quad \mbox{and}\quad B=\{\max_{0 \le i \le \tau^*} \left | \frac{\hat{\mathcal{J}}_{n_0+i, M, H}}{\hat{\sigma}_{{n_0, M, H}}} \right | > a\}. 
\]

Second, the probability on the right hand side of (\ref{ine}) is
\begin{eqnarray}
\mbox{P}_{0}\{T^*< t \}&=&\mbox{P}_{0} \biggl(\max_{0 \le i \le t} \left | \frac{\hat{\mathcal{J}}_{n_0+i, M, H}}{\hat{\sigma}_{{n_0, M, H}}} \right | > a \biggr)\nonumber\\
&=&\mbox{P}_{\tau} \biggl(\max_{\tau^* \le i \le t+\tau^*} \left | \frac{\hat{\mathcal{J}}_{n_0+i, M, H}}{\hat{\sigma}_{{n_0, M, H}}} \right | > a \biggr). \label{eq1}
\end{eqnarray}
The last equation holds because both probabilities are based on the observations after the change points $0$ and $\tau-n_0$, respectively, and the observations have the same distribution. From (\ref{eq1}), we define the event
\[
C=\{\max_{\tau^* \le i \le t+\tau^*} \left | \frac{\hat{\mathcal{J}}_{n_0+i, M, H}}{\hat{\sigma}_{{n_0, M, H}}} \right | > a \}.
\] 

From the above defined events $A, B$ and $C$, we see that $A=B \cup C$. Therefore, $\mbox{P}(A)=\mbox{P}(B)+\mbox{P}(C)-\mbox{P}(B\cap C)$. From the definition of the events $B$ and $C$, we see that if $B$ occurs, then $T^*<0$ or the stopping time $T_H(a, M)<\tau^*$. Then $C$ cannot occur. Therefore $\mbox{P}(A)=\mbox{P}(B)+\mbox{P}(C)$. Moreover, from the definitions of $A, B, C$, (\ref{eq2}) becomes
\[
\mbox{P}_{\tau}(T^*<t| T^*>0)=\frac{\mbox{P}_{\tau}(A)-\mbox{P}_{\tau}(B) }{1-\mbox{P}_{\tau}(B)}=\frac{\mbox{P}_{\tau}(C)}{1-\mbox{P}_{\tau}(B)}=\frac{\mbox{P}_{0}\{T^*< t \}}{1-\mbox{P}_{\tau}(B)},  
\]
where the last equation holds by using (\ref{eq1}). Then (\ref{ine}) can be proved accordingly. This completes the proof that the supremum of the EDDs attains at the immediate change point $\tau=n_0$.  

We next establish the upper bound for the EDDs. To simplify notation, we let $\hat{\mathcal{J}}_{T}\equiv \hat{\mathcal{J}}_{n,M,H}$, which is the test statistic evaluated at the stopping time $T_H(a, M)$.
Let $y_{i,r_1r_2} = x_{i,r_1}x_{i,r_2}$, where $x_{i,r_1}$ and $x_{i,r_2}$ are the $r_1$th and $r_2$th component of $X_i$, respectively. Hence, $\E(y_{i,r_1r_2}) = \mbox{Cov}(x_{i,r_1},x_{i,r_2}) = \sigma_{i,r_1r_2},$ which is the $r_1$th row and $r_2$th column of $\Sigma_i$.
Using (\ref{estimator}), we see that
\bda
\E(\hat{\mathcal{J}}_T)
&=&\E\bigg\{\frac{1}{H^2}\sum_{r_1,r_2=1}^p\sum_{i,j=1}^{H} W_M(i,j)y_{T+i,r_1r_2}y_{T+j,r_1r_2}\bigg\}\\
&=&\E\bigg\{\frac{1}{H^2}\sum_{r_1,r_2=1}^p\sum_{i,j=1}^{H} W_M(i,j)\sigma_{T+i,r_1r_2}\sigma_{T+j,r_1r_2}\bigg\}\\
&+&\E\bigg\{\frac{1}{H^2}\sum_{r_1,r_2=1}^p\sum_{i,j=1}^{H} W_M(i,j)(y_{T+i,r_1r_2}-\sigma_{T+i,r_1r_2})(y_{T+j,r_1r_2}-\sigma_{T+j,r_1r_2})\bigg\}\\
&+&\E\bigg\{\frac{2}{H^2}\sum_{r_1,r_2=1}^p\sum_{i,j=1}^{H} W_M(i,j)\sigma_{T+i,r_1r_2}(y_{T+j,r_1r_2}-\sigma_{T+j,r_1r_2})\bigg\}\\
&=&\E(I)+\E(II)+\E(III).
\eda
By Lemma 3, 4, and 5 in the supplementary material, we obtain
$$\E(I)\geq\frac{1}{H}\E\{(T-M)(T-M-1)\}{\rm tr}\{(\Sigma_{\tau}-\Sigma_{\tau+1})^2\}\{1+o(1)\},$$
$$\E(II)=O\Big[\log(H)\mbox{tr} \{(\Sigma_{\tau}-\Sigma_{\tau+1})^2\}\Big]=o\Big[\sqrt{H}\mbox{tr}\{(\Sigma_{\tau}-\Sigma_{\tau+1})^2\}\Big],$$
and
$$\E(III)=o\Big[\sqrt{H}\mbox{tr}\{(\Sigma_{\tau}-\Sigma_{\tau+1})^2\}\Big].$$
Hence, 
\bea
\E(\hat{\mathcal{J}}_{T})
&\geq&\frac{1}{H}\E\{(T-M)(T-M-1)\}{\rm tr}\{(\Sigma_{\tau}-\Sigma_{\tau+1})^2\}\{1+o(1)\}\nonumber\\
&+&o\Big[\sqrt{H}\mbox{tr}\{(\Sigma_{\tau}-\Sigma_{\tau+1})^2\}\Big]. \label{expectation}
\eea

Using (\ref{expectation}), we have
\begin{eqnarray}
a\cdot \sigma_{H,M,0}&\geq&\frac{1}{H}\E\{(T-M)(T-M-1)\}{\rm tr}\{(\Sigma_{\tau}-\Sigma_{\tau+1})^2\}\{1+o(1)\}\nonumber\\
&-&(|\E(\hat{\mathcal{J}}_{T})|-a \cdot \sigma_{H,M,0})
+o\Big[\sqrt{H}\mbox{tr}\{(\Sigma_{\tau}-\Sigma_{\tau+1})^2\}\Big].\label{ub-s}
\end{eqnarray}

Let $\hat{\mathcal{J}}_{T-1}$ denote the test statistic evaluated at $T-1$. From the stopping rule (\ref{stopping-rule}), we have
$$
\E|\hat{\mathcal{J}}_{T-1}|\leq a\cdot \sigma_{H,M,0}.
$$
By Jensen's inequality and triangle inequality, we also have
\bda
\E|\hat{\mathcal{J}}_{T-1}|
&\geq&|\E(\hat{\mathcal{J}}_{T})|-|\E(\hat{\mathcal{J}}_{T}-\hat{\mathcal{J}}_{T-1})|.
\eda
Combining the above two inequality, we obtain
\be
|\E(\hat{\mathcal{J}}_{T})|-a\cdot \sigma_{H,M,0} \leq|\E(\hat{\mathcal{J}}_{T}-\hat{\mathcal{J}}_{T-1})|. \label{ub_1}
\ee

Based on similar derivations,
%$$\E(\hat{\mathcal{J}}_{T-1})=\frac{1}{H}\E\{(T-M-1)(T-M-2)\}{\rm tr}\{(\Sigma_{\tau}-\Sigma_{\tau+1})^2\}\{1+o(1)\},$$
%which leads to the result that
\bea
|\E(\hat{\mathcal{J}}_{T}-\hat{\mathcal{J}}_{T-1})|
&=&\frac{2}{H}\E(T-M-1){\rm tr}\{(\Sigma_{\tau}-\Sigma_{\tau+1})^2\}\{1+o(1)\}\nonumber\\
&+&o\Big[\sqrt{H}\mbox{tr}\{(\Sigma_{\tau}-\Sigma_{\tau+1})^2\}\Big].\label{ub_2}
\eea
Combining (\ref{ub-s}), (\ref{ub_1}) and (\ref{ub_2}), we obtain
$$\frac{1}{H}\E\{(T-M-2)^2\}{\rm tr}\{(\Sigma_{\tau}-\Sigma_{\tau+1})^2\}\{1+o(1)\}\le a\cdot \sigma_{H,M,0}+o\Big[\sqrt{H}\mbox{tr}\{(\Sigma_{\tau}-\Sigma_{\tau+1})^2\}\Big].$$

Using $a\cdot \sigma_{H,M,0}=O\{H^{r}\cdot\mbox{tr}{(\Sigma_{\tau}-\Sigma_{\tau+1})^2}\}$ with ${1}/{2} \le r <1$ and the Jensen's inequality, we have
$$
\E(T-M-2)\leq\sqrt{\E\{(T-M-2)^2\}} \le \bigg[\frac{a\cdot \sigma_{H,M,0}\cdot H}{\mbox{tr}\{(\Sigma_{\tau}-\Sigma_{\tau+1})^2\}}\bigg]^{1/2}\{1+o(1)\}.
$$
This completes the proof of Theorem 2.

\subsection{Proof of Theorem 3}

The asymptotic normality of $\hat{\mathcal{J}}_{n_0, M}$ can be established by the martingale central limit theorem. Toward this end, we let
$\mathscr{F}_0=\{\emptyset, \Omega \}$, $\mathscr{F}_k=\sigma\{X_1,...,X_k\}$ with $k=1,2,...,n_0$, and $\mbox{E}_k(\cdot)$ denote the
conditional expectation given $\mathscr{F}_k$. Define
$D_{n_0,k}=(\mbox{E}_k-\mbox{E}_{k-1})\hat{\mathcal{J}}_{n_0, M}$ and it is easy to see that $\hat{\mathcal{J}}_{n_0, M}-\mu_{\hat{\mathcal{J}}_{n_0, M}}=\sum_{k=1}^{n_0}D_{n_0,k}$.

We further define $S_{n_0,m}=\sum_{k=1}^{m}D_{n_0,k}=\mbox{E}_m\hat{\mathcal{J}}_{n_0, M}-\mu_{\hat{\mathcal{J}}_{n_0, M}}$. We can show that
for $q \ge m$, $\mbox{E}(S_{n_0,q}|\mathscr{F}_m)=S_{n_0,m}$. To this end, we note that
$S_{n_0,q}=\mbox{E}_q\hat{\mathcal{J}}_{n_0, M}-\mu_{\hat{\mathcal{J}}_{n_0, M}}=\mbox{E}_m\hat{\mathcal{J}}_{n_0, M}-\mu_{\hat{\mathcal{J}}_{n_0, M}}+\mbox{E}_q\hat{\mathcal{J}}_{n_0, M}
-\mbox{E}_m\hat{\mathcal{J}}_{n_0, M}=S_{n_0,m}+(\mbox{E}_q\hat{\mathcal{J}}_{n_0, M}
-\mbox{E}_m\hat{\mathcal{J}}_{n_0, M})$. Then
\begin{eqnarray}
\mbox{E}(S_{n_0,q}|\mathscr{F}_m)&=&S_{n_0,m}+\mbox{E}\{\mbox{E}_q(\hat{\mathcal{J}}_{n_0, M})|\mathscr{F}_m\}
-\mbox{E}\{\mbox{E}_m(\hat{\mathcal{J}}_{n_0, M})|\mathscr{F}_m\}\nonumber\\
&=&S_{n_0,m}+\mbox{E}\{\mbox{E}_m(\hat{\mathcal{J}}_{n_0, M})\}-\mbox{E}\{\mbox{E}_m(\hat{\mathcal{J}}_{n_0, M})\}\nonumber\\
&=&S_{n_0,m}.\nonumber
\end{eqnarray}

As a result, we see that $\{S_{n_0,k}, \mathscr{F}_k\}$ is a martingale and accordingly,  $\{D_{n_0,k}, 1\le k \le n_0\}$ is a
martingale difference sequence with respect to the $\sigma$-fields
$\{\mathscr{F}_k,1\le k \le n_0\}$

Based on similar derivations for Lemmas 2 and 3 in Li and Chen (2012), we can show that under (\ref{model}) and (C1)-(C2), as $n_0 \to \infty$,
\begin{eqnarray}
\frac{\sum_{k=1}^{n_0}\mbox{E}_{k-1}(D_{n_0,k}^2)}{\sigma^2_{{\hat{\mathcal{J}}_{n_0, M}}}}\xrightarrow{p} 1. \nonumber
\end{eqnarray}
And,
\begin{eqnarray}
\frac{\sum_{k=1}^{n_0}\mbox{E}(D_{n_0,k}^4)}{\sigma^4_{{\hat{\mathcal{J}}_{n_0, M}}}}\to
0.\nonumber
\end{eqnarray}
The above two results are sufficient conditions for the martingale central limit theorem. This thus completes the first part of Theorem 3.

To show the second part of Theorem 3, we only need to show the ratio consistency of $\hat{\sigma}_{\hat{\mathcal{J}}_{n_0, M},0}$ defined in (\ref{var.est2}) to ${\sigma}_{\hat{\mathcal{J}}_{n_0, M},0}$ under the null hypothesis.
From the expression (\ref{var.est}), we apply (\ref{model}) such that under the null hypothesis,
\begin{eqnarray}
\mbox{E}\biggl(\frac{1}{n^*}\sum_{s,t}^*X_{t+h_2}^T X_s X_{s+h_1}^T X_t\biggr)
&=&\frac{1}{n^*}\sum_{s,t}^* \mbox{E}(Z^T \Gamma_{t+h_2}^T\Gamma_s Z Z^T \Gamma_{s+h_1}^T\Gamma_t Z )\nonumber\\
&=&{\mbox{tr}\{C(h_1)C(h_2)\}}.\nonumber
\end{eqnarray}
This shows that $\mbox{E}[\widehat{\mbox{tr}\{C(h_1)C(h_2)\}}]={\mbox{tr}\{C(h_1)C(h_2)\}}$. Similarly, under the conditions (C1)-(C2), we have $\mbox{Var}[\widehat{\mbox{tr}\{C(h_1)C(h_2)\}}]=o[{\mbox{tr}^2\{C(h_1)C(h_2)\}}]$. This implies that under the null hypothesis, $$\widehat{\mbox{tr}\{C(h_1)C(h_2)\}}/ \mbox{tr}\{C(h_1)C(h_2)\}\xrightarrow{p} 1.$$
The second part of Theorem 3 is then proved by applying the continuous mapping theorem.

\subsection{Proof of Theorem 4}

We first show that $\mbox{P}(\hat{M}=M)=1$ as $n_0 \to \infty$. Note that the event that $\hat{M} > M$ is equivalent to the event that 
\[
\widehat{\mbox{tr}\{C(M+1)C(-M-1)\}}/ \widehat{\mbox{tr}\{C(0)C(0)\}} > \epsilon. 
\]  
Therefore, $\mbox{P}(\hat{M}>M)$ is equivalent to 
\[
\mbox{P}\biggl[\widehat{\mbox{tr}\{C(M+1)C(-M-1)\}}/ \widehat{\mbox{tr}\{C(0)C(0)\}} > \epsilon\biggr]. 
\]  
It is also equivalent to $\mbox{P}\biggl[\widehat{\mbox{tr}\{C(M+1)C(-M-1)\}}/ {\mbox{tr}\{C(0)C(0)\}} > \epsilon\biggr]$ as $$\widehat{\mbox{tr}\{C(0)C(0)\}}/ \mbox{tr}\{C(0)C(0)\}\xrightarrow{p} 1$$ from the proof of Theorem 3.  

From (\ref{var.est}), we can show that $\mbox{E}[\widehat{\mbox{tr}\{C(M+1)C(-M-1)\}}]=0$ and $$\mbox{Var}\biggl[\widehat{\mbox{tr}\{C(M+1)C(-M-1)\}}/ {\mbox{tr}\{C(0)C(0)\}}\biggr]= O(n^{-2}).$$ 
Using Chebyshev's inequality, we can show that  as $n_0 \to \infty$, 
$$\mbox{P}\biggl[\widehat{\mbox{tr}\{C(M+1)C(-M-1)\}}/ {\mbox{tr}\{C(0)C(0)\}} > \epsilon\biggr]=0, $$ 
or equivalently, $\mbox{P}(\hat{M}>M)=0$. Similarly, we can show that $\mbox{P}(\hat{M}<M)=0$. We then establish the consistency of $\hat{M}$ to $M$.

To prove $\mbox{E}_{\infty}\{T_H(a, \hat{M})\} =\mbox{E}_{\infty}\{T_H(a, M)\}$, we only need to show that as $n_0 \to \infty$, 
\[
\mbox{P}_{\infty}\{T_H(a, \hat{M}) \le t\} =\mbox{P}_{\infty}\{T_H(a, M) \le t\}.  
\]
Toward this end, we notice that
\begin{eqnarray}
\mbox{P}_{\infty}\{T_H(a, \hat{M})\le t\}&=&\mbox{P}_{\infty}\{T_H(a, \hat{M})\le t, \hat{M}=M\}\nonumber\\
&+& \mbox{P}_{\infty}\{T_H(a, \hat{M})\le t, \hat{M}\ne M\}, \nonumber
\end{eqnarray}
where the second term converges to zero because $\mbox{P}(\hat{M}=M)=1$ as $n_0 \to \infty$. 

To prove $\mbox{E}_{0}\{T_H(a, \hat{M})\} =\mbox{E}_{0}\{T_H(a, M)\}$, we notice that as $n_0 \to \infty$, 
\begin{eqnarray}
\mbox{E}_{0}\{T_H(a, \hat{M})\} &=& \mbox{E}(\mbox{E}_{0}\{T_H(a, \hat{M})|\hat{M}\})\nonumber\\
&=&\mbox{E}_{0}\{T_H(a, M)\} \mbox{P}(\hat{M}=M)\nonumber\\
&+&\sum_{M^{*}\ne M} \mbox{E}_{0}\{T_H(a, M^{*})\} \mbox{P}(\hat{M}=M^*)\nonumber\\
&=&\mbox{E}_{0}\{T_H(a, M)\}.\nonumber
 \end{eqnarray}
This completes the proof of Theorem 4.

\end{document}